\documentclass[11pt]{article}

\usepackage{amssymb,amsmath,amsthm,amsfonts}
\usepackage{graphicx}
\usepackage[usenames]{color}
\usepackage{algorithm}
\usepackage{algpseudocode}
\usepackage{url}

\graphicspath{{./pics/}}

\newtheorem{rmk}{Remark}[section]

\newtheorem{exam}{Example}

\def \O {\Omega}
\def \veps {\varepsilon}

 \def\p{\partial} \def\nb{\nonumber}
\def \Vh0{\stackrel{\circ}{V}_h} 
   
\def\Om{\Omega}   
\newcommand{\q}{\quad}    
\def\l{\label}    

\def\m{\mbox}   
\def\ms{\medskip}  \def\ss{\smallskip}

\def\u{{\bf u}}  
  \def\x{{\bf x}} 
 \def\H{{\bf H}} 

\newcommand{\lc}
{\mathrel{\raise2pt\hbox{${\mathop<\limits_{\raise1pt\hbox
{\mbox{$\sim$}}}}$}}}

\newcommand{\gc}
{\mathrel{\raise2pt\hbox{${\mathop>\limits_{\raise1pt\hbox{\mbox{$\sim$}}}}$}}}

\newcommand{\ec}
{\mathrel{\raise2pt\hbox{${\mathop=\limits_{\raise1pt\hbox{\mbox{$\sim$}}}}$}}}

\def\bb{\begin{equation}} \def\ee{\end{equation}}

\def\beqn{\begin{eqnarray}}  \def\eqn{\end{eqnarray}}

\def\beqnx{\begin{eqnarray*}} \def\eqnx{\end{eqnarray*}}

\def\bn{\begin{enumerate}} \def\en{\end{enumerate}}

\def\bd{\begin{description}} \def\ed{\end{description}}

\title{A New Splitting Method
for Time-dependent Convection-dominated Diffusion
Problems}

\author{Feng Shi\thanks{Shenzhen Institutes
of Advanced Technology, Chinese Academy of Sciences, Shenzhen
518055,  China.  The work of this author was partially supported by NSFC (Project No.
41104039). ({\tt feng.shi@siat.ac.cn}). }
        \and Guoping Liang\thanks{Beijing
FEGEN Software Co., Ltd. ({\tt guopingliang@yahoo.com.cn
 }).}
        \and Yubo Zhao\thanks{Shenzhen Institutes
of Advanced Technology, Chinese Academy of Sciences, Shenzhen
518055, China. The work of this author was partially supported by
the Knowledge Innovation Program of the Chinese Academy of Sciences
(China) under KJCX2-EW-L01, and the international cooperation
project of Guangdong provience (China) under 2011B050400037. ({\tt
yb.zhao@siat.ac.cn}).} \and Jun Zou
\thanks{Corresponding author. Department of Mathematics, The Chinese University
of Hong Kong, Shatin, NT, Hong Kong. ({\tt zou@math.cuhk.edu.hk})}
 }

\begin{document}
\maketitle

\begin{abstract}
We present a new splitting method for time-dependent
convection-dominated diffusion problems. The original convection
diffusion system is split into two sub-systems: a pure convection
system and a diffusion system. At each time step,  a convection
problem and a diffusion problem are solved successively.
The scheme has the following nice features:
the convection subproblem is solved explicitly and a multistep technique
is introduced to essentially enlarge the stability region
so that the resulting scheme behaves like an unconditionally stable scheme;
the diffusion subproblem is always self-adjoint and coercive so that
it can be solved efficiently using many existing optimal
preconditioned iterative solvers.  The scheme is then extended
for Navier-Stokes equations, where the nonlinear convection is
resolved by a linear explicit multistep scheme at the convection
step, and only a  generalized Stokes problem is needed to solve
at the diffusion step with the resulting stiffness matrix being invariant
in the time marching process.
The new schemes are all free from tuning some stabilization
parameters for the convection-dominated diffusion problems.
Numerical simulations are presented to
demonstrate the stability, convergence and performance of the
single-step and multistep variants of the new scheme.

\end{abstract}




\pagestyle{myheadings} \thispagestyle{plain} \markboth{A new
splitting method}{Feng Shi, Guoping Liang, Yubo Zhao and Jun Zou}

\section{Introduction}

In this work we shall propose a new fully discrete splitting scheme
for solving the convection-dominated diffusion problems of the following general form
\begin{equation}\label{gcdcd}
  u_t+\nabla\cdot(\mathbf{b} u)-\nabla\cdot(\veps \nabla u)+cu=F  \quad \mbox{in}\quad \Om\times(0,T)
\end{equation}
with the boundary and initial conditions
\begin{equation}\label{gcdcd_bc}
u= u_b \quad \mbox{on}\quad \p\Om\times(0,T)\,; \quad
u(0,{\bf x})=u_0({\bf x}) \quad \mbox{in}\quad \Om
\end{equation}
where $\Om$ is an open bounded polyhedral domain in $\mathbb{R}^d$
($d=1,2,3$) with boundary $\Gamma=\p\Omega$, and $T$ is the
terminal time. Functions $\mathbf{b}$, $\veps$  and $c$ in (\ref{gcdcd}) are
the convective field,  diffusion and reactive coefficients respectively,
while $F$, $u_b$ and
$u_0$ are the source term, the boundary and initial data
respectively. As we are mainly interested in the construction of
numerical schemes, we will not specify detailed regularity
conditions on all these coefficients to ensure the well-posedness of
the initial-boundary value problem (\ref{gcdcd})-(\ref{gcdcd_bc}).
%

The new fully discrete splitting scheme is then extended for Navier-Stokes equations
\begin{equation}\label{ns}
\left\{
\begin{aligned}
  \mathbf{u}_t+(\mathbf{u}\cdot\nabla) \mathbf{u}-Re^{-1}\triangle \mathbf{u}+\nabla p&=\mathbf{F} \quad \mbox{in}\quad \Om\times (0,T)\\
\nabla\cdot\mathbf{u}&=0 \quad \mbox{in}\quad \O\times(0,T)
\end{aligned}\right.
\end{equation}
with the boundary and initial conditions
\begin{equation}\label{ns_bc}
\u= \u_b \quad \mbox{on}\quad \p\Om\times(0,T)\,; \quad
\u(0,{\bf x})=\u_0({\bf x}) \quad \mbox{in}\quad \Om
\end{equation}
where  $\mathbf{u}$, $p$, $\mathbf{F}$ and  $Re$ are respectively
the velocity, pressure, body force and Reynolds number,
while $\mathbf{u}_b$ and $\u_0$ are the given boundary and initial
data.

The numerical solution of a time-dependent problem requires
a discretization in both time and space, and some linearization
if the concerned problem is nonlinear. A great variety of time marching schemes are
available in the literature, such as the classical methods like
the forward and backward Euler schemes, the Crank-Nicolson scheme,
the Adams-Bashforth method etc. Operator splitting is also a popular technique
for time discretization, such as the Yanenko method, the Peaceman-Rachford method,
the Douglas-Rachford method and the $\theta$ scheme; see \cite{QV1994, DH2003, G1989}
and references therein.

In solving the convection-dominated diffusion equations and the
Navier-Stokes equations with large Reynolds numbers, it is
known that standard finite element methods perform poorly and may
exhibit nonphysical oscillations. Many spatial stabilization
techniques have been proposed and studied. The streamline-upwind
Petrov-Galerkin method was developed
for convective transport problems \cite{HB1979,BH1982}, and its
basic idea is to modify the standard Petrov-Galerkin formulation by
adding a streamline upwind perturbation, which acts only in the flow
direction and is solely defined in the interiors of elements. The
Galerkin least-squares method \cite{HFH1989} is a conceptual
simplification of the streamline-upwind Petrov-Galerkin method, and
adds a stabilization that involves an element-by-element weighted
least-squares of the residual to the original differential equation.
The efficiency of these two stabilization techniques may be affected by
the choices of the stabilization parameters involved. There are still no
precise general formulae to help  select optimal parameters in
numerical simulations; see,  e.g., \cite[Remark\,10.4]{S2005}. These
stabilization parameters may depend possibly also on time steps
for time-dependent problems, so their choices become more tricky in
practice as we have to balance between temporal and spatial errors
\cite{JN2011}.

By changing the sign of the convective term in the weighted
least-squares formulation, the unusual stabilized finite element
method (USFEM) can achieve the absolute stability for any positive
stabilization parameter involved in the scheme, but it is still a
tricky and inconclusive technical issue of how to choose this
parameter in order to obtain good accuracy
\cite{FFH1992,FF1992,FF1995,FV2000}. The variational multiscale
method was developed based on the inherent multiscale structure of
solutions \cite{H1995,HFMQ1998,HMJ2000,JKL2006}. This method
defines the large scales by a projection into an appropriate
subspace, but may also involve the technical issue of how to select a
stabilization parameter to balance the stability and accuracy.

As it is known \cite{BH1982}, explicit Galerkin solutions for flow problems
could be quite under-diffusive, effectively increasing the Peclet or Reynolds number.
Furthermore, explicit methods are generally conditionally stable.
But explicit schemes have the advantages that
they may not need to solve systems of algebraic equations \cite{CT1985}
or the resulting stiffness matrices stay the same in the time marching process.

The characteristic-based-split (CBS) method has been widely studied
for fluid and solid dynamic problems
\cite{ZC1995a,ZC1995b,ZNCVO1999,NCZ2006}, and we refer to the
monograph \cite{ZTN2005} and the references therein for its detailed
introduction and various applications. This method is based on the
splitting of the convection and diffusion parts. The convection part
is formally handled by the standard characteristics method, where
the numerical solutions at the current time are updated by the
approximations at the previous time. But the schemes need to locate
some spatial points based on the characteristics, and the spatial points
are likely no longer grid points of the spatial discretization. One way to avoid this
is to adjust the meshes, while another way is to apply the standard interpolation
to evaluate the values of the solutions at these spatial points using the values of the solutions
and other quantities at grid points.
An alternative technique, used in the CBS method, is to approximate
numerical solutions at computed spatial points by the solutions and
other quantities at grid points by Taylor expansion. In addition,
the CBS method needs to approximate the average convective field,
for which different treatments may lead to different schemes, such
as fully explicit, semi-implicit or implicit ones, and also
different stabilization effects \cite{NCZ2006,ZTN2005}.

In the derivation of the new scheme, we shall use the same operator splitting
as the CBS method did, to split the convection diffusion system into
a purely convective part and a diffusive part.
The diffusion part is  discretized by the standard backward scheme.
But the central difference from the CBS method lies in our new treatment
of the convection part, which is completely independent of the characteristic curves
and any spatial grid points used, unlike the CBS method.

Another novel idea of the new method is the flexibility in its
special explicit treatment of the convection part: we can
recursively execute the explicit convection step up to a finite
number of times with smaller local time steps  during one
diffusion correction. This can essentially improve the stability of
the resulting scheme so it may behave
like an unconditionally stable scheme.

The rest of the paper is arranged as follows. The single-step scheme is first derived
for the convection diffusion equation in Section\,\ref{sec:gcdcd}, and its multistep variant
in Section\,\ref{sec:m_teps}. The new scheme is then extended in
Section\,\ref{sec:ns} for the Navier-Stokes equations.
Numerical experiments are carried out in Section\,\ref{sec_numer_test}
to check the accuracy, stability and performance of the new schemes,
as well as to investigate how the stability condition can be improved
by the multistep scheme compared with the single-step one.
At the end of this numerical section, the driven cavity flow problem is tested
with the new scheme and compared with the benchmark results to demonstrate
the validity of the new method. Some concluding remarks are given
in Section\,\ref{sec_conclus}.

\section{Derivation of algorithms}\label{sec:algorithm}
In this section we derive a new method for solving the
convection-dominated diffusion equation (\ref{gcdcd}). For the purpose
we introduce some notations. We first partition the time interval $[0,
T]$: $0=t_0<t_1<\cdots <t_N=T$, with $t_n=n\Delta t$ and $\Delta
t=T/N$. We will use $u^n$ and $u^{n+\frac12}$ respectively for
the approximate values of $u(\cdot, t)$ at $t=t_n$ and
$t_{n}+{\Delta t}/{2}$. But when $u(\cdot, t)$ is a known function, $u^n$
and $u^{n+\frac12}$ will stand for its exact values at $t=t_n$ and
$t_{n}+{\Delta t}/{2}$, e.g., $f^n=f(\cdot,t_n)$, and ${\bf
b}^n={\bf b}(\cdot,t_n)$.

\subsection{Single-step scheme for the convection diffusion
equation}\label{sec:gcdcd} We first adopt the standard operator
splitting technique \cite{G1989} and split the convection diffusion
equation (\ref{gcdcd}) into a pure convection equation and a
diffusion equation. Then we approximate two equations in time by
the central difference and backward Euler schemes respectively  to
obtain
         \begin{eqnarray}
            \dfrac{u_*^{n+1}-u^n}{\Delta t}+\nabla\cdot(\mathbf{ b}^{n+\frac12} \,u^{n+\frac12})
            &=&{  f}^{\,n+\frac12},
            \label{convect_1}\\
            \dfrac{u^{n+1}-u_*^{n+1}}{\Delta t}-\nabla\cdot(\veps \nabla
            u^{n+1})+c^{n+1}u^{n+1}&=&g^{n+1},\label{diffu_1}
         \end{eqnarray}
where $f$ and $g$ can be any functions such that $F=f+g$. However
in order to have a unified principle for the selection of the components
$f$ and $g$ for both the convection diffusion equation and
Navier-Stokes equations, we will suggest some special choice
of $f$ and $g$ later on; see Remark\,\ref{rm_source_f_0}. 


We shall use finite element methods to solve (\ref{convect_1}) and (\ref{diffu_1}) respectively
for the solutions $u_*^{n+1}$ and $u^{n+1}$. To do so, we need the variational
formulations of these two equations. For equation (\ref{diffu_1}), it is straightforward to derive
its variational form:

Find $u^{n+1}\in H^1(\Om)$ such that $u^{n+1}=u_b^{n+1}$ on ${\Gamma}$ and solves
   \begin{equation}\label{varfor_diffu_1}
            (u^{n+1},v)+\Delta t(\veps \nabla u^{n+1},\nabla v)+\Delta t(c^{n+1}u^{n+1},v)=(u_*^{n+1},v)
            +\Delta t( g^{n+1},v) \q \forall\,v\in H^1_0(\Om)\,.
         \end{equation}

On the other hand, the solution of the convection step
(\ref{convect_1}) is more tricky. Clearly the scheme is implicit and
involves the solution of a linear convection equation. The main idea
of this work is to propose an explicit scheme to solve this linear
convection equation. For this aim, we apply the Taylor's expansion to
compute $u^{n+\frac12}$ by the values at previous times, and can write
\begin{equation*}
    u^{n+\frac12}\approx u({\bf x},t_n+\frac{\Delta t}{2})=u({\bf x},t_n)+\frac{\Delta t}{2}u_t({\bf x},t_n)+O(\Delta t^2),
\end{equation*}
then using the convection equation
\begin{equation}\label{convection}
    u_t+\nabla\cdot(\mathbf{b} \,u)= f
\end{equation}
we deduce
\begin{equation}\label{taylor_1}
    u^{n+\frac12}\approx u^n+\frac{\Delta t}{2}\left(f^n-\nabla\cdot(\mathbf{b}^n u^n)\right)=:\xi^n\,.
\end{equation}
Using this relation, we can rewrite (\ref{convect_1}) as
         \begin{equation}\label{m_convect_2}
            \dfrac{u_*^{n+1}-u^n}{\Delta t}+\nabla\cdot\left(\mathbf{ b}^{n+\frac12} \xi^n\right)={
            f}^{\,n+\frac12}.
         \end{equation}

Noting that (\ref{convection}) is a pure convective equation, only partial
boundary condition on the inflow boundary should be imposed, namely
         \begin{equation}\label{cd_inflow_bdy}
\Gamma_t^-:=\{\mathbf{x}\in\Gamma; ~
\mathbf{b}(\mathbf{x}, t)\cdot\mathbf{n}({\bf x})<0\}\,
         \end{equation}
where $\mathbf{n}({\bf x})$ is the outward normal to the boundary of $\Om$ at
$\x$.
Accordingly we should set a similar condition on the inflow boundary
associated with the scheme (\ref{m_convect_2}). So for any positive integer $n$,
we define
         \begin{equation}\label{cd_inflow_bdy2}
\Gamma_{n}^-:=\{\mathbf{x}\in\Gamma; ~
\mathbf{b}^{n}(\x) \cdot\mathbf{n}({\bf x})<0\}\,.
         \end{equation}
As the exact solution is specified on the entire boundary (cf.\,(\ref{gcdcd})),
it is natural for us to assume the values for the solution $u_{*}^{n+1}$ to (\ref{m_convect_2})
on the inflow boundary $\Gamma_{n+1}^-$:
          \begin{equation}
u_{*}^{n+1}=u_b^{n+1}\, \quad \mbox{on} \quad  \Gamma_{n+1}^-\,.\label{art_bc}
         \end{equation}
This induces the following test space for the scheme (\ref{m_convect_2}):
$$
H^1_{\Gamma_{n+1}^-}(\Om)=\left\{w\in
H^1(\Om); ~w=0 ~~\mbox{on} ~~{\Gamma_{n+1}^-}\right\}\,.
$$
Now multiplying a test function $v\in H^1_{\Gamma_{n+1}^-}(\Om)$ on
both sides of (\ref{m_convect_2}), and integrating over $\Om$ and
using the integration by parts we obtain \beqn
(u_*^{n+1},v)&=&(u^{n},v)+\Delta t ({ f}^{\,n+\frac12},v)\nb\\
   &&+\, \Delta t(\xi^n,{\bf { b}}^{n+\frac12}\cdot\nabla v)-\Delta t<\xi^n,v{\bf { b}}^{n+\frac12}\cdot {\bf n}>_{\Gamma\setminus\Gamma_{n+1}^-}\nb\\
       &=&(u^{n},v) +\Delta t ({ f}^{\,n+\frac12},v)\nb\\
    &&+\, \Delta t\left(u^n+\frac{\Delta t}{2}\left(f^n- \nabla\cdot
(\mathbf{b}^n u^n)\right), {\bf { b}}^{n+\frac12}\cdot\nabla v\right)\,  \l{convect_1_variational_form}\\
&&-\, \Delta t\langle u^n+\frac{\Delta t}{2}\left(f^n-\nabla\cdot
(\mathbf{b}^n u^n)\right), v{\bf { b}}^{n+\frac12}\cdot {\bf
n}\rangle_{\Gamma\setminus\Gamma_{n+1}^-}.\nb \eqn

%
It remains to introduce the spatial discretizations for both
equations (\ref{varfor_diffu_1}) and (\ref{convect_1_variational_form}), which we will do
by finite element methods.
Assume that $V_h$ is a finite element space approximating the
Sobolev space $H^1(\Om)$, and $I_h$ is the interpolation operator of
$H^1(\Om)$ into $V_h$. Then based on the variational formulations
(\ref{convect_1_variational_form}) and (\ref{varfor_diffu_1}),
 we propose the following single-step scheme for solving the convection-dominated diffusion
 problem (\ref{gcdcd}).

\ms
{\bf  Algorithm 1 (Single-step scheme).}
\begin{description}
  \item[Step 0.] Compute the initial value $u_h^0=I_hu_0$. For each $n=0, 1, \cdots, N-1$, do the
  following.
    \item[Step 1.] Find $u_{h,*}^{n+1}\in V_h$ such that
    $u_{h,*}^{n+1}=I_hu_b^{n+1}$ on $\Gamma_{n+1}^-$ and it solves
         \begin{equation*}\label{alg_A_1}
         \begin{split}
         &
         {\,\ } \hskip-1truecm
         (u_{h,*}^{n+1},v_h)=(u_h^{n},v_h) +\Delta t ({ f}^{\,n+\frac12},v_h)\\
&+\Delta t\Big(u_h^n+\frac{\Delta
t}{2}\left(f^n-  \nabla \cdot (\mathbf{b}^n u_h^n)\Big), \,
{\bf { b}}^{n+\frac12}\cdot\nabla v_h\right)\\
    &-\Delta t\Big\langle u_h^n+\frac{\Delta
    t}{2}\Big(f^n- \nabla \cdot (\mathbf{b}^n u_h^n)\Big), \,v_h{\bf { b}}^{n+\frac12}\cdot
{\bf n}\Big\rangle_{\Gamma\setminus\Gamma_{n+1}^-} ~\forall\, v_h\in V_h \cap
H^1_{\Gamma_{n+1}^-}(\Om)\,.
         \end{split}
         \end{equation*}
  \item[Step 2.] Find $u_h^{n+1}\in V_h$ such that $u_h^{n+1}=I_hu_b^{n+1}$ on $\Gamma$
  and it solves
  $$
     {\,\ } \hskip-1truecm
     (u_h^{n+1},v_h)+\Delta t(\veps \nabla u_h^{n+1},\nabla v_h)+\Delta t(c^{n+1}u_h^{n+1},v_h)
            =(u_{h,*}^{n+1},v_h)+\Delta t( g^{n+1},v_h) \q \forall\, v_h\in V_h\cap H_0^1(\Om)\,.
  $$
 \end{description}

\begin{rmk}\label{lumped_rmk}
One may compute the term $(u_{h,*}^{n+1},v_h)$ in Step 1 by
the standard mass-lumping technique \cite{CT1985},  then $u_{h,*}^{n+1}$ can
be computed explicitly without solving a linear system.
\end{rmk}

\subsection{Multistep scheme for the convection diffusion
equation}\label{sec:m_teps}

The focus of this work is mainly on the case when
the convection diffusion system (\ref{gcdcd}) is convection-dominated.
For this case the stability of the explicit single-step scheme (Algorithm 1)
may pose severe restrictions on time steps, leading to sufficiently small
time steps and great computational efforts for the entire numerical resolution
process.

To improve the stability, we may
execute the convection step (Step 1) a few times for each diffusion
correction (Step 2) so that we can use much smaller time steps
for the convection part and much larger time steps for the diffusion step.
To do so, we write the result $u_{h,*}^{n+1}$ of Step 1 formally as
\begin{eqnarray}
u_{h,*}^{n+1} &=& F_{conv}^{CD}\Big(\Delta t, f^{n}, f^{n+1}, {\bf
b}^n, {\bf b}^{n+1}, u_h^n, u_b^{n+1}\Big)\,. \label{appr_u_1}
\end{eqnarray}
Then the multistep scheme is to run this convection step $m$ times
with smaller time step size ${\Delta t}/m$ for $u_{h,*}^{n+1}$, namely we compute
\begin{eqnarray}
u_{h,*}^{n+\frac{i}{m}} &=& F_{conv}^{CD}\Big(\frac{\Delta t}{m},
f^{n+\frac{i-1}{m}}, f^{n+\frac{i}{m}},  {\bf b}^{n+\frac{i-1}{m}},
{\bf b}^{n+\frac{i}{m}}, u_{h,*}^{n+\frac{i-1}{m}},
u_b^{n+\frac{i}{m}}\Big), \label{appr_u_multistep_1}
\end{eqnarray}
recursively  for $i=1,2,\cdots, m$, with  $u_{h,*}^n=u_h^n$.

We shall  call $\delta t={\Delta t}/m$ and $\Delta t$  as the
local time step size and the global time step size respectively.
Replacing Step 1 by the multistep iteration
(\ref{appr_u_multistep_1}), we propose the following multistep
scheme for the convection diffusion equation (\ref{gcdcd}).

\ms {\bf  Algorithm 2 (Multistep scheme with index $m$).}
  \begin{description}
 \item[Step 0.] Compute the initial value $u_h^0=I_hu_0$. For each $n=0, 1, \cdots, N-1$, do the
  following.
  \item[Step 1.]  Set $u_{h,*}^{n}= u_h^{n}$. For $i=1,2, \cdots, m$,
  compute $u_{h,*}^{n+\frac{i}{m}}\in V_h$
  such that $u_{h,*}^{n+\frac{i}{m}}=I_hu_b^{n+\frac{i}{m}}$ on $\Gamma_{n+{i}/{m}}^-$
  and it solves for all $v_h\in V_h \cap H^1_{\Gamma_{n+i/m}^-}(\Om)$,
  \begin{eqnarray*}
      && {\,\ } \hskip-2.5truecm
       (u_{h,*}^{n+\frac{i}{m}},v_h)=
         (u_{h,*}^{n+\frac{i-1}{m}},v) +\delta t ( f^{n+\frac{2i-1}{2m}},v_h)\nb\\
&& {\, } \hskip-2truecm + \delta t\left(u_{h,*}^{n+\frac{i-1}{m}}+
\frac{\delta t}2 (f^{n+\frac{i-1}{m}}  -  \nabla\cdot
(\mathbf{b}^{n+\frac{i-1}{m}} u_{h,*}^{n+\frac{i-1}{m}})),
{\bf b}^{n+\frac{2i-1}{2m}}\cdot\nabla  v_h\right)\nb\\
    && {\, } \hskip-2truecm -\delta t\left<u_{h,*}^{n+\frac{i-1}{m}}
    +\frac{\delta t}2\left( f^{n+\frac{i-1}{m}} - \nabla\cdot (\mathbf{b}^{n+\frac{i-1}{m}}
u_{h,*}^{n+\frac{i-1}{m}})\right),  v_h{\bf
b}^{n+\frac{2i-1}{2m}}\cdot {\bf
n}\right>_{\Gamma\setminus{\Gamma_{n+{i}/{m}}^-}}.
         \end{eqnarray*}
  \item[Step 2.] Compute $u_h^{n+1}\in V_h$ such that $u_h^{n+1}=I_hu_b^{n+1}$ on $\Gamma$
  and it solves for all $v_h\in V_h\cap H_0^1(\Om)$,
     \begin{equation*}
            (u_h^{n+1},v_h)+\Delta t(\veps \nabla u_h^{n+1},\nabla v_h)+\Delta t(c^{n+1}u_h^{n+1},v_h)=(u_{h,*}^{n+1},v_h)+\Delta t( g^{n+1},v_h)\,.
         \end{equation*}
\end{description}


\section{Single-step and multistep schemes for Navier-Stokes equations}\label{sec:ns}
We are now going to extend the new schemes proposed in Sections\,\ref{sec:gcdcd}-\ref{sec:m_teps} for the convection-dominated diffusion equation
to the Navier-Stokes equations (\ref{ns}). For the purpose, we split
the system (\ref{ns}) into a pure convection system and
a diffusion system (the generalized Stokes problem) as follows:
         \begin{eqnarray}
            \dfrac{\mathbf{u}_*^{n+1}-\mathbf{u}^{n}}{\Delta t}+(\mathbf{u}^{n+\frac12}\cdot\nabla)
            \mathbf{u}^{n+\frac12}&=&\mathbf{ f}^{n+\frac12},
            \label{NS_convect_1}\\
 \dfrac{\mathbf{u}^{n+1}-\mathbf{u}_*^{n+1}}{\Delta t}-Re^{-1} \triangle
            \mathbf{u}^{n+1}+\nabla p^{n+1}&=&\mathbf{g}^{n+1}, \label{NS_diffu_1}\\
            \nabla \cdot \bf{u}^{n+1}&=&0\,.  \label{NS_div}
         \end{eqnarray}

It is straightforward to derive the variational form of
the generalized Stokes system (\ref{NS_diffu_1})-(\ref{NS_div}):

Find ${\bf u}^{n+1}\in \H^1(\Om)$ and $p\in L_0^2(\Om)$ such that
${\bf u}^{n+1}={\bf u}_b^{n+1}$ on ${\Gamma}$ and it solves
\begin{eqnarray}
{\ } \hskip-1truecm (\Delta
t)^{-1}(\mathbf{u}^{n+1},\mathbf{v})+Re^{-1}( \nabla
\mathbf{u}^{n+1}, \nabla \mathbf{v})
            -(p^{n+1},\nabla\cdot \mathbf{v})&=&(\Delta t)^{-1}(\mathbf{u}_{*}^{n+1},\mathbf{v})+(
            \mathbf{g}^{n+1},\mathbf{v})\,, \label{varfor_ns_diffu_1}\\
(\nabla\cdot\mathbf{u}^{n+1},q)&=&0  \label{varfor_ns_div} 
\end{eqnarray}
for any $\mathbf{v} \in \H_0^1(\Om) $ and $q\in L_0^2(\Om)$.

Next we will do the same as we did in Section\,\ref{sec:gcdcd} to
propose an explicit scheme for solving the convection system
(\ref{NS_convect_1}). To do so, we first handle the nonlinear convection term
involving $\mathbf{u}^{n+\frac12}$. In fact, combining the Taylor's
expansion
\begin{equation*}
    {\bf u}^{n+\frac12}\approx {\bf u}({\bf x},t_n+\frac{\Delta t}{2})={\bf u}({\bf x},t_n)+\frac{\Delta t}{2}{\bf u}_t({\bf x},t_n)+O(\Delta t^2),
\end{equation*}
and the pure convection equation
\begin{equation}\label{NS_convection}
    {\bf u}_t+(\mathbf{u} \cdot\nabla){\bf u}= {\bf f},
\end{equation}
we can obtain a similar approximation to (\ref{taylor_1}) but in a vector-valued form:
\begin{equation}\label{NS_taylor_1}
    {\bf u}^{n+\frac12}\approx {\bf u}^n+\frac{\Delta t}{2}\left({\bf f}^n-({\bf u}^n \cdot\nabla){\bf u}^n\right)=:\eta^n.
\end{equation}

Again, we introduce the inflow boundary
$$\Gamma_{n+1}^-=\left\{{\bf x}\in \Om;\, {\bf u}_b^{n+1}\cdot {\bf n}({\bf x})<0\right\}.$$
Then we can write by using integration by parts  for
any $\,{\bf v}\in \mathbf{H}^1(\Om)$ with ${\bf
v}|_{\Gamma_{n+1}^-}=0$ that
\begin{equation}\label{NS_convect_ibp}
\left(( \eta^n\cdot\nabla) \eta^n,{\bf v}\right)=\left< \eta^n,
 \eta^n\cdot\mathbf{n}\,\mathbf{v}\right>_{\Gamma\setminus\Gamma_{n+1}^-}
 -\left(\eta^n,\nabla\cdot \eta^n{\bf v}\right)-\left( \eta^n,( \eta^n\cdot\nabla){\bf v} \right),
\end{equation}
using this relation and plugging (\ref{NS_taylor_1}) in (\ref{NS_convect_1}) we derive
the variational form of (\ref{NS_convect_1}):
\beqn
(\mathbf{u}_*^{n+1},\mathbf{v})&=&(\mathbf{u}^{n},\mathbf{v})+\Delta t (\mathbf{f}^{n+\frac{1}{2}},\mathbf{v})
         -\Delta t\Big< \eta^n,  (\eta^n\cdot\mathbf{n}) \mathbf{v}\Big>_{\Gamma\setminus\Gamma_{n+1}^-}\nb\\
        && +\Delta t\Big( \eta^n, (\nabla\cdot\eta^n  +  \eta^n\cdot\nabla)  {\bf v}\Big)\,. \label{NS_convect_vf_1}
\eqn

\begin{rmk}\label{rm_source_f_0}
We observe from the formulation (\ref{NS_convect_vf_1}) that
$\nabla\cdot {\bf f}^n$ is needed in the term $\nabla\cdot{\eta}^n$,
hence it adds some extra regularity on the source component ${\bf
f}$. This suggests us to better choose $\bf {f} \equiv0$ in the
decomposition ${\bf F}={\bf f}+{\bf g}$ for the Navier-Stokes
equations so that the new scheme does not need the evaluation of
$\nabla\cdot {\bf f}^n$, unlike in the CBS methods
\cite{ZNCVO1999,NCZ2006,ZTN2005}. For the unification of the numerical
schemes for both the convection diffusion equation and Navier-Stokes
equations, we shall always select $f\equiv 0$ in Algorithms 1 and 2
from now on for the convection diffusion equation.
\end{rmk}


Let ${\bf V}_h$ and $M_h$ be two finite element spaces approximating
the Sobolev space ${\bf H}^1(\Om)$ and $L_0^2(\Om)$, and ${\bf I}_h$
be a interpolation operator of ${\bf H}^1(\Om)$ into ${\bf V}_h$. By
virtue of the variational formulations (\ref{NS_convect_vf_1}) and
(\ref{varfor_ns_diffu_1}), we propose a single-step scheme for solving the
Navier-Stokes equations (\ref{ns}).

\ms {\bf  Algorithm 3 (Single-step scheme).}

    \begin{description}
  \item[Step 0.] Compute the initial value ${\bf u}_h^0={\bf I}_h{\bf u}_0$. For each $n=0, 1, \cdots, N-1$, do the
  following.
    \item[Step 1.]
 Find ${\bf u}_{h,*}^{n+1}\in {\bf V}_h$ such that
    ${\bf u}_{h,*}^{n+1}={\bf I}_h{\bf u}_b^{n+1}$ on $\Gamma_{n+1}^-$ and it solves
 \beqnx
 (\mathbf{u}_{h,*}^{n+1},\mathbf{v})&=&(\mathbf{u}_{h}^{n},\mathbf{v}_h)
    -\Delta t\Big<\eta_{h}^{n}, (\eta_{h}^{n}\cdot\mathbf{n})\mathbf{v}_h\Big>_{\Gamma\setminus\Gamma_{n+1}^-}\\
&&+\Delta t\Big(\eta_{h}^{n}, (\nabla\cdot\eta_{h}^{n} + \eta_{h}^{n}\cdot\nabla){\bf
v}_h\Big)
\q \forall\, {\bf v}_h\in {\bf V}_h \cap {\bf
H}^1_{\Gamma_{n+1}^-}(\Om)\,.
\eqnx

  \item[Step 2.] Find ${\bf u}_h^{n+1}\in {\bf V}_h$ and $p_h\in M_h$, such that ${\bf u}_h^{n+1}={\bf I}_h{\bf u}_b^{n+1}$ on $\Gamma$
  and it solves
  \begin{eqnarray*} {\,\ } \hskip-1truecm
\Delta t^{-1}(\mathbf{u}_h^{n+1},\mathbf{v}_h)+Re^{-1}(\nabla
\mathbf{u}_h^{n+1},\nabla \mathbf{v}_h)
            -(p_h^{n+1},\nabla\cdot \mathbf{v}_h)&=&\Delta t^{-1}(\mathbf{u}_{h,*}^{n+1},\mathbf{v}_h)
            +(\mathbf{g}^{n+1},\mathbf{v}_h),\\
(\nabla\cdot \mathbf{u}_h^{n+1},q_h)&=&0
\end{eqnarray*}
${\, }$ \hskip0.48truecm for any $\mathbf{v}_h\in
\mathbf{V}_h\cap {\bf H}^1_0(\Om)$ and
$q_h\in M_h$.
 \end{description}

For simplicity we have used in Algorithm 3 the notation $\eta_{h}^{n}$,
which is defined as $\eta^{n}$ in (\ref{NS_taylor_1}) but with $\u^n$
replaced by $\u_h^n$. Similarly we shall use the following notation in
Algorithm 4:
$$
  \eta_{h,*}^{n+\frac{i-1}{m}}={\bf
         u}_{h,*}^{n+\frac{i-1}{m}}-\frac{\Delta t}{2}({\bf u}_{h,*}^{n+\frac{i-1}{m}}
\cdot\nabla){\bf u}_{h,*}^{n+\frac{i-1}{m}}\,.
$$

We observe from Algorithm 3 that the nonlinear convection term
$(\mathbf{u}\cdot\nabla)\mathbf{u}$ in Navier-Stokes equations has
been  treated explicitly in the time marching process, which may
severely restrict the time step size in order to ensure the
stability of the scheme when the convection is dominated in comparison with
the diffusion of the flow system.
To improve the stability, we may apply Step
1 several times with a smaller time step size during one diffusion
correction (Step 2). For this purpose we write the result of Step 1
formally as
\begin{eqnarray}
{\bf u}_{h,*}^{n+1} &=& {\bf F}_{conv}^{NS}\Big(\Delta t, {\bf
u}_h^n, {\bf u}_b^{n+1}\Big)\,. \label{appr_u_2}
 \end{eqnarray}
Then a multistep variant of this scheme is to execute this step
$m$ times with a smaller time step size ${\Delta t}/m$ to derive
$\u_{h,*}^{n+1}$:
\begin{eqnarray}
\u_{h,*}^{n+\frac{i}{m}} &=& {\bf F}_{conv}^{NS}\Big(\frac{\Delta
t}{m}, {\bf u}_{h,*}^{n+\frac{i-1}{m}}, {\bf
u}_b^{n+\frac{i}{m}}\Big)\, \label{appr_u_multistep_2}
\end{eqnarray}
 for $i=1,2,\cdots, m$, with  ${\bf u}_{h,*}^n={\bf u}_h^n$.
 This leads to the following multistep scheme for the Navier-Stokes
equations.

\ms {\bf  Algorithm 4 (Multistep scheme with index $m$).}

 \begin{description}
 \item[Step 0.] Compute the initial value $\mathbf{u}_h^0=\mathbf{I}_h\mathbf{u}_0$.
 For each $n=0, 1, \cdots, N-1$, do the
  following.
  \item[Step 1.]  Set ${\bf u}_{h,*}^{n}={\bf u}_h^{n}$;
  ~~then for $i=1,2, \cdots, m$,

  compute $\mathbf{u}_{h,*}^{n+\frac{i}{m}}\in \mathbf{V}_h$
  such that  $\mathbf{u}_{h,*}^{n+\frac{i}{m}}=\mathbf{I}_h\mathbf{u}_b^{n+\frac{i}{m}}$ on $\Gamma_{n+{i}/{m}}^-$
  and it solves
  \begin{eqnarray*}
 (\mathbf{u}_{h,*}^{n+\frac{i}{m}},\mathbf{v}_h)&=&(\mathbf{u}_{h,*}^{n+\frac{i-1}{m}},\mathbf{v}_h)
    -\delta t\left<\eta_{h,*}^{n+\frac{i-1}{m}}, (\eta_{h,*}^{n+\frac{i-1}{m}}\cdot\mathbf{n})\mathbf{v}_h\right>_{\Gamma\setminus\Gamma_{n+i/m}^-}\\
&& +\delta
t\left(\eta_{h,*}^{n+\frac{i-1}{m}},  (\nabla\cdot\eta_{h,*}^{n+\frac{i-1}{m}}
+ \eta_{h,*}^{n+\frac{i-1}{m}}\cdot\nabla)  {\bf v}_h\right)
\q \forall\,{\bf v}_h\in {\bf V}_h \cap {\bf
H}^1_{\Gamma_{n+i/m}^-}(\Om)\,.
         \end{eqnarray*}
  \item[Step 2.] Compute $(\mathbf{u}_h^{n+1},p_h^{n+1})\in \mathbf{V}_h\times M_h$ such that $\mathbf{u}_h^{n+1}=\mathbf{I}_h\mathbf{u}_b^{n+1}$ on $\Gamma$
  and it solves
\begin{eqnarray*}
 {\,\ } \hskip-1truecm
\Delta t^{-1}(\mathbf{u}_h^{n+1},\mathbf{v}_h)+Re^{-1}(\nabla
\mathbf{u}_h^{n+1},\nabla \mathbf{v}_h)
            -(p_h^{n+1},\nabla\cdot \mathbf{v}_h)&=&\Delta t^{-1}(\mathbf{u}_{h,*}^{n+1},\mathbf{v}_h)+(
            \mathbf{g}^{n+1},\mathbf{v}_h),\\
(\nabla\cdot \mathbf{u}_h^{n+1},q_h)&=&0
\end{eqnarray*}
${\, }$ \hskip0.48truecm  for any $(\mathbf{v}_h,q_h)\in
(\mathbf{V}_h\cap {\bf H}^1_0(\Om))\times M_h$.
\end{description}

\begin{rmk}
The second steps in Algorithms 3 and 4 can be replaced by the
projection-type methods so that the pair of finite element spaces
for approximating the velocity and pressure does not need to meet
the LBB condition and only  Poisson problems are needed to solve for
updating both the velocity and pressure. For the projection method,
we refer to the pioneering work by Chorin \cite{C1968} and Temam
\cite{T1969}.
\end{rmk}

\section{Numerical experiments}\label{sec_numer_test}

In this section we shall carry out two sets of numerical tests to
check the actual convergence orders of the single-step and multistep
schemes proposed in the previous two sections and how the multistep
scheme improves the stability of the single-step scheme.

Let ${\cal T}_h$ be a regular triangulation of $\Om$, with
$h_K=\mbox{diam}(K)$ for $K\in {\cal T}_h$, and $h=\max_{K\in {\cal T}_h}{h_K}$.
We shall use the following linear finite element space $V_h\subset H^1(\Om)$:
\begin{equation}\label{eq:linear}
V_h=\{w_h\in H^1(\Om); \,w_h|_{K}\in P_1(K) \q \forall K\in {\cal
T}_h\}
\end{equation}
for the solution of the convection diffusion equation (\ref{gcdcd}), and the following Taylor-Hood
finite element spaces \cite{TH1973}
\begin{eqnarray}
\mathbf{V}_h&=&\{\mathbf{v}_{h}\in H^1(\Om)^2;
\,\mathbf{v}_h|_{K}\in P_2(K)^2 \q \forall K\in {\cal T}_h\}, \\
M_h&=&\{q_{h}\in H^1(\Om); \, q_h|_{K}\in P_1(K) \q \forall K\in
{\cal T}_h\}\, \label{Taylor-Hood_element}
\end{eqnarray}
for the solution of the Navier-Stokes Equations (\ref{ns}).

We recall that we have used the central finite difference scheme for the convection
diffusion equation and the backward Euler scheme for the diffusion equation
in time discretization. Therefore it is natural for us to expect the following
numerical convergence orders when the finite element spaces in (\ref{eq:linear})
and (\ref{Taylor-Hood_element}) are used:
$$
\|{ u}^N -{u}_h^N\|_{L^2(\O)} \le C (h^2+\Delta t)
$$
for the convection diffusion equation (\ref{gcdcd}),  and
$$ \|{\bf u}^N -{\bf u}_h^N\|_{L^2(\O)} \le C (h^3+\Delta t)
\q \m{and} \q  \|p^N-p_h^N\|_{L^2(\O)} \le C (h^2+\Delta t)
$$
respectively for the velocity and pressure of the Navier-Stokes equations (\ref{ns}).
Naturally we may think that the convergence of the scheme can be improved
by using a second-order scheme for the diffusion equation in time discretization, but
our numerical experiments have firmly disapproved this conjecture.
We are currently investigating the possible treatments to help construct the schemes
which have second order temporal accuracy.

We remark that all the errors shown in this section are the $L^2$-norm errors at the terminal
time $t=T$ unless specified otherwise.

\subsection{Tests for the convection diffusion equation}\label{subsec_numer_test_cd}

We first apply the new single-step and multistep schemes to the
following two examples which are taken from references \cite{JN2011}
and \cite{JKL2006}.

\begin{exam}\label{exa3}
The coefficients and domain in equation (\ref{gcdcd}) are taken to be
the following:
$$
d=2, \q T=1,  \q \veps=10^{-8}, \q \mathbf{b}=(1,-1)^T, \q c=1\,, \q
\Om=(0,1)^2
$$
with the exact solution given by
$
u(x, y, t)=e^{2\pi t}\sin(2\pi x)\sin(2\pi y).
$
\end{exam}
This example is a slight modification of the one in \cite{JN2011},
where $e^{\sin(2\pi t)}$ is used. Instead we use $e^{2\pi t}$, which
makes the solution vary in a much larger range, namely in the
interval $[-e^{2\pi},e^{2\pi}]$, and has a much larger norm,
$\|u(\cdot,1)\|=\frac12e^{2\pi}\approx267.7458$.

\begin{exam}\label{exa1}
The coefficients and domain in equation (\ref{gcdcd}) are taken to be
the following:
$$
d=2, \q T=1,  \q \veps=10^{-8}, \q \mathbf{b}=(2,-1)^T, \q c=1\,, \q
\Om=(0,1)^2
$$
with the exact solution given by
$
u(x, y, t)=t^2\cos(xy^2)\,.
$
\end{exam}

To compute the actual convergence orders of the numerical schemes,
we shall use the uniform triangulations of domain $\Om$ with
triangular elements in all our numerical simulations.

\subsubsection{Convergence Tests for the single-step
scheme}\label{subsec_conver_test_cd}

In order to find the actual convergence order of the single-step scheme (Algorithm 1) in time,
we choose a very small mesh size and then observe the changes of the errors when
the time step size is halved. Similarly
we will do the other way around when we try to find  the actual convergence order of the single-step
scheme (Algorithm 1) in space.

Tables 1 and 2 show the $L^2$-norm errors with different mesh sizes
when the time step size is fixed for Examples 1 and 2 respectively.
Clearly we see the second order spatial convergence of the
single-step scheme (Algorithm 1).

\begin{table}[htbp]
\tabcolsep 0pt\caption{Convergence results of Algorithm 1 for
Example \ref{exa3} with fixed $\Delta t= 1/2^{16}$}\vspace*{-10pt}
\begin{center} \def\temptablewidth{\textwidth}
{\rule{\temptablewidth}{1pt}}
\begin{tabular*}{\temptablewidth}{@{\extracolsep{\fill}}llc}
 $h$ & $\|u-u_h\|$ & order\\ \hline
$1/4$ & 9.18526(+1) &-\\
 $1/8$ & 1.84780(+1) &2.3135\\
$1/16$& 4.24054 &2.1235\\
 $1/32$& 1.03466&2.0351\\
 $1/64$& 2.54797(-1) &2.0217\\
 $1/128$&6.36669(-2) &2.0007\\ \hline
\end{tabular*}
\end{center}\label{tas}
\end{table}

\begin{table}[htbp]
\tabcolsep 0pt\caption{Convergence results of Algorithm 1 for
Example \ref{exa1} with fixed $\Delta t= 1/2^{16}$}\vspace*{-10pt}
\begin{center} \def\temptablewidth{\textwidth}
{\rule{\temptablewidth}{1pt}}
\begin{tabular*}{\temptablewidth}{@{\extracolsep{\fill}}llc}

 $h$ & $\|u-u_h\|$ & order\\ \hline
 $1/4$ & 9.76826(-3)&-\\
$1/8$ &2.41756(-3)&2.0145\\
$1/16$& 6.02478(-4) &2.0046\\
$1/32$& 1.49729(-4) &2.0086\\
$1/64$&3.69132(-5) &2.0201 \\
 $1/128$& 8.70186(-6) &1.9687 \\ \hline\end{tabular*}
\end{center}\label{tbs}
\end{table}

Now we fix the uniform mesh size at $h=1/128$, and run the single-step scheme
(Algorithm 1) for Examples \ref{exa3} and \ref{exa1} with
the following sequence of time step sizes
\begin{equation}\label{cd_time_step_series}
   \Delta t=0.1/2^k\,,  \q k=-1,0,1, 2, \cdots
\end{equation}
to find out the stability region of the numerical scheme. The
numerical results are listed in Tables \ref{tat} and \ref{tbt}, from
which we observe that Algorithm 1 does not converge till $k=6$ and
$7$ respectively for Examples \ref{exa3} and \ref{exa1},
corresponding to two rather small time step sizes of $\Delta
t=1/640$ and $1/1280$. Such restrictions on time step sizes are
natural, required by the stability condition for the explicit time
marching scheme we have used for the convection step in Algorithm 1.
As we shall see in the next
subsection, the new multistep scheme can essentially improve the
stability condition.

  \begin{table}[htbp]
\tabcolsep 0pt\caption{Convergence results of Algorithm 1 for
Example \ref{exa3} with fixed $h=1/128$}\vspace*{-10pt}
\begin{center} \def\temptablewidth{\textwidth}
{\rule{\temptablewidth}{1pt}}
\begin{tabular*}{\temptablewidth}{@{\extracolsep{\fill}}llc}

 $\Delta t$ & $\|u-u_h\|$ & order\\
 \hline
$0.1/2^5$ & divergence &-\\
$0.1/2^6$& 2.02151 &-\\
$0.1/2^7$&1.01181 &0.9985\\
\hline
\end{tabular*}
\end{center}\label{tat}
\end{table}

 \begin{table}[htbp]
\tabcolsep 0pt\caption{Convergence results of Algorithm 1 for
Example \ref{exa1} with fixed $h=1/128$}\vspace*{-10pt}
\begin{center} \def\temptablewidth{\textwidth}
{\rule{\temptablewidth}{1pt}}
\begin{tabular*}{\temptablewidth}{@{\extracolsep{\fill}}llc}

 $\Delta t$ & $\|u-u_h\|$ & order\\
 \hline
$0.1/2^6$&divergence &-\\
$0.1/2^7$&1.71484(-4) &-\\
\hline
\end{tabular*}
\end{center}\label{tbt}
\end{table}

%
%
%
%

%

\subsubsection{Stability improvement by the multistep scheme}\label{subsec_multi_step_test_cd}
We can observe from the previous subsection that the single-step
scheme (Algorithm 1) may provide the expected convergence and
preserve the accurate convergence orders when it converges. However,
this scheme requires sufficiently small time step size as shown in
Tables \ref{tat} and \ref{tbt}, hence may restrict its applications
in practice. The multistep scheme (Algorithm 2) is proposed to
improve the stability of the single-step scheme. This section is to
test how the multistep scheme can improve the stability region.

We note that $\Delta t$ is the global time step size, which is used
for the diffusion correction. As we are interested mainly in the
convection-dominated diffusion problems, the time step size required
for the convection is usually much smaller than the one for the
diffusion.

In our numerical tests, for each fixed global time step $\Delta t=0.1/2^k$ ($k=-1, 0, 1, 2, \cdots$),
we run the multistep scheme with index $m=1, 2^1, 2^2, \cdots$ until we observe
the convergence of the scheme, and then record the corresponding index $m$;
see Tables \ref{tab_cd_t1_vm} and \ref{tab_cd_t2_vm} for the recorded index $m$
corresponding to each fixed $\Delta t$ and the resulting relative $L^2$-norm error
of the approximate solution.

As we see from Table \ref{tab_cd_t1_vm}, when we take $\Delta t =0.1$,
which is too large for the stability of  the explicit scheme involved in the convection step,
but we can still achieve the convergence of the multistep scheme
with index $m\ge 64$.
Tables \ref{tab_cd_t1_vm} and \ref{tab_cd_t2_vm} have demonstrated that
though the single-step scheme does not converge for a fixed global time step $\Delta t$,
the multistep scheme always converges when the index $m$ is appropriately large.
So we can conclude that if we take an appropriately large index $m$, say $m=30$,
the multistep scheme can be viewed as an unconditionally stable scheme.

Furthermore, we have also computed the convergence orders of the
multistep scheme in terms of the global time step size for Examples
1 and 2 with a fixed index $m$ and mesh size $h$. The results are
shown in Tables \ref{tab_cd_t1_m_64} and \ref{tab_cd_t2_m_128}.
Combining these results with the ones for the single-step scheme
(cf.\,Table \ref{tat}), we can clearly observe the first order
temporal convergence for both examples.

\begin{table}[htbp]
\tabcolsep 0pt\caption{Stability of Algorithm 2 for
Example \ref{exa3} with index $m$ and fixed $h=1/128$} \vspace*{-10pt}
\begin{center} \def\temptablewidth{\textwidth}
{\rule{\temptablewidth}{1pt}}
\begin{tabular*}{\temptablewidth}{@{\extracolsep{\fill}}lcl}

 $\Delta t$ & $m$  & $\frac{\|{u}-{u}_h\|}{\|{u}\|}$  \\
\hline
$0.1/2^6$&1& 7.55011(-3) \\
$0.1/2^5$&2& 1.54379(-2)   \\
$0.1/2^4$&4& 3.15569(-2) \\
$0.1/2^3$&8&  6.41671(-2)  \\
$0.1/2^2$&16& 1.30693(-1)  \\
$0.1/2^1$ &32&  2.69788(-1)  \\
$0.1$& 64&   5.68483(-1)  \\
$0.2$&128&   1.26837 \\
 \hline
\end{tabular*}
\end{center}\label{tab_cd_t1_vm}
\end{table}

\begin{table}[htbp]
\tabcolsep 0pt\caption{Stability of  Algorithm 2 for Example \ref{exa1}
with index $m$ and fixed $h=1/128$}\vspace*{-10pt}
\begin{center} \def\temptablewidth{\textwidth}
{\rule{\temptablewidth}{1pt}}
\begin{tabular*}{\temptablewidth}{@{\extracolsep{\fill}}lrl}

 $\Delta t$ &m & $\frac{\|{u}-{u}_h\|}{\|{u}\|}$ \\
\hline
$0.1/2^7$&1&1.65924(-4)\\
$0.1/2^6$&2& 7.46950(-4) \\
$0.1/2^5$&4& 1.96066(-3)   \\
$0.1/2^4$&8& 4.39337(-3) \\
$0.1/2^3$&16& 9.28328(-3)  \\
$0.1/2^2$&32&  1.96240(-2)  \\
$0.1/2^1$ &64& 4.11452(-2)  \\
$0.1$& 128& 8.41251(-2)  \\
$0.2$&256& 1.70957(-1) \\
\hline
\end{tabular*}
\end{center}\label{tab_cd_t2_vm}
\end{table}

\begin{table}[htbp]
\tabcolsep 0pt\caption{Convergence order of Algorithm 2 for
Example \ref{exa3} with fixed index $m=64$ and $h=\frac 1{128}$
 }\vspace*{-10pt}
\begin{center} \def\temptablewidth{\textwidth}
{\rule{\temptablewidth}{1pt}}
\begin{tabular*}{\temptablewidth}{@{\extracolsep{\fill}}lll}

 $\Delta t$  & $\|u-u_h\|$&order  \\
\hline
$0.1$&   1.52209(+2)&-  \\
$0.1/2^1$ &  7.23099(+1)&1.0738  \\
$0.1/2^2$&3.51127(+1)  &1.0422\\
$0.1/2^3$&  1.73718(+1) &1.0152 \\
$0.1/2^4$& 8.66679 &1.0032\\
$0.1/2^5$& 4.34772&0.9952 \\
$0.1/2^6$& 2.18766& 0.9909\\
\hline
\end{tabular*}
\end{center}\label{tab_cd_t1_m_64}
\end{table}

\begin{table}[htbp]
\tabcolsep 0pt\caption{Convergence order of Algorithm 2 for
Example \ref{exa1} with fixed index $m=128$ and $h=\frac 1{128}$}\vspace*{-10pt}
\begin{center} \def\temptablewidth{\textwidth}
{\rule{\temptablewidth}{1pt}}
\begin{tabular*}{\temptablewidth}{@{\extracolsep{\fill}}llc}

 $\Delta t$ & $\|u-u_h\|$ &order \\
\hline

$0.1$& 8.69440(-2) &- \\
$0.1/2^1$ & 4.27948(-2) &1.0227 \\
$0.1/2^2$&2.07747(-2)  &1.0426\\
$0.1/2^3$& 1.00067(-2)&1.0538  \\
$0.1/2^4$& 5.01775(-3) &0.9959\\
$0.1/2^5$& 2.54242(-3)  & 0.9808\\
$0.1/2^6$& 1.31725(-3) & 0.9487\\
\hline
\end{tabular*}
\end{center}\label{tab_cd_t2_m_128}
\end{table}

Next, we carry out some numerical tests to check how the multistep
scheme can improve the stability region quantitatively. For each
fixed mesh size $h$, we increase the index $m$ gradually and record
the largest global time step size $\Delta t$ that can ensure the
convergence of the entire algorithm. And the largest time step size
will be written as the critical time step size $\Delta t_{crit}$ for
the stability of the algorithm.  The results are shown in Tables
\ref{tab_cd_t1_critical_time_step} and
\ref{tab_cd_t2_critical_time_step}, from which we can see that the
stability region is nearly doubled when the index $m$ of the
multistep scheme is doubled. So the multistep scheme can indeed
clearly and essentially enlarge the stability of the entire
algorithm.


\begin{table}[htbp]
\tabcolsep 0pt\caption{Critical global time step size $\Delta
t_{crit}$ of  Algorithm 2 for Example \ref{exa3} in terms of index
$m$}\vspace*{-10pt}
\begin{center} \def\temptablewidth{\textwidth}
{\rule{\temptablewidth}{1pt}}
\begin{tabular*}{\temptablewidth}{@{\extracolsep{\fill}}lcccccc}

 $m$ & 1 & 2& 10& 20& 40&80 \\
 \hline
 $h=1/64$\\
 $\Delta t_{crit}$&
0.0049&0.0093&0.046&0.093&0.18&0.37\\

\hline
$h=1/128$\\
 $\Delta t_{crit}$&
0.0024&0.0045&0.022&0.045&0.091&0.18\\
\hline
\end{tabular*}
\end{center}\label{tab_cd_t1_critical_time_step}
\end{table}

\begin{table}[htbp]
\tabcolsep 0pt\caption{Critical global time step size $\Delta
t_{crit}$ of Algorithm 2 for Example \ref{exa1}  in terms of index
$m$}\vspace*{-10pt}
\begin{center} \def\temptablewidth{\textwidth}
{\rule{\temptablewidth}{1pt}}
\begin{tabular*}{\temptablewidth}{@{\extracolsep{\fill}}lcccccc}

 $m$ & 1 & 2& 10& 20& 40&80 \\
 \hline
 $h=1/64$\\
 $\Delta t_{crit}$&
0.0032&0.0060&0.029&0.058&0.11&0.23\\

\hline
$h=1/128$\\
 $\Delta t_{crit}$&
0.0015&0.0030&0.014&0.028&0.057&0.11\\
\hline
\end{tabular*}
\end{center}\label{tab_cd_t2_critical_time_step}
\end{table}

We remark that we have done many more numerical experiments for
Examples 1 and 2, but with the diffusion coefficients $\veps$
varying in a wider range, from $10^{-3}$ to $10^{-15}$, and many
different convective vectors ${\bf b}$, and observed similar
convergence and stability behaviors for the single-step and
multistep schemes  as we have shown above.

\subsection{Tests for the Navier-Stokes Equations}\label{subsec_conver_test_ns}

Now we will apply our new single-step and multistep schemes
(Algorithms 3 and 4) to two examples of Navier-Stokes equations with
analytical solutions to check the actual convergence orders of the
schemes and how the multistep scheme improves the stability
of the single-step scheme. Then we will apply these schemes to the
benchmark problem of the lid-driven cavity flow to verify their
validity.

\begin{exam}\label{ex_ns_1}
Consider the Navier-Stokes equations (\ref{ns}) with the following
parameters:
$$
\Omega= [0,1]^2, \q T=1, \q
Re=5000 \q \m{and} \q 10000
$$
with the exact solution $(\mathbf{u},p)=(u_1,u_2,p)$ given by
$p=(x^2-y^2)\cos(t)$ and
\begin{equation*}
u_1=10x^2(x-1)^2y(y-1)(2y-1)\cos(t)\,, \q
u_2=-10x(x-1)(2x-1)y^2(y-1)^2\cos(t)\,.
\end{equation*}
\end{exam}

\begin{exam}\label{ex_ns_2}
Consider the Navier-Stokes equations (\ref{ns}) with the same
parameters as in Example\,\ref{ex_ns_1}, but the exact solution
$(\mathbf{u},p)=(u_1,u_2,p)$ given by
\begin{equation*}
u_1=t^3y^2\,, \q
u_2=t^2x\,, \q
p=tx+y-(t+1)/2\,.
\end{equation*}
This is an example where only a discretization error in time
occurs \cite{JMR2006}.
\end{exam}

\subsubsection{Convergence Tests for the single-step scheme}\label{subsec_conver_test_ns_single}

We first verify the convergence orders of the single-step scheme (Algorithm 3) in both
space and time for Example \ref{ex_ns_1}.
Tables \ref{tab_ns_t1_1}-\ref{tab_ns_t1_2} present
the convergence results in time for the Reynolds numbers $Re=5000$ and $10000$ respectively, with
a fixed uniform mesh of size $h=1/128$, and
Tables \ref{tab_ns_h1}-\ref{tab_ns_h2}
give the convergence results in space for the Reynolds numbers $Re=5000$ and $10000$ respectively,
with a fixed $\Delta t=10^{-6}$.
From these tables we can clearly see  the optimal first order  convergence
of the single-step scheme in time and
the optimal third and second order convergence in space
respectively for the velocity and pressure.

For Example \ref{ex_ns_2}, we have tested the single-step scheme
(Algorithm 3) with the Reynolds numbers $Re=5000$ and $10000$, and
the uniform mesh of size $h=1/48$ and $1/64$, and the sequence of
time step sizes as listed in (\ref{cd_time_step_series}). The
results have shown that the scheme converges only when the time step
size $\Delta t=0.1/2^k$ is sufficiently small, namely when $k$ takes
at least $4$ ($\Delta t=1/160$) and $5$ ($\Delta t=1/320$)
respectively for $h=1/48$ and $1/64$. This test indicates that the
single-step scheme may require sufficiently small time step size to
ensure its convergence, as one can expect for this strongly convection-dominated
example. In the next
Section\,\ref{subsec_multistep_test_ns} we will show the multistep
scheme (Algorithm 4) can essentially improve the stability of the
single-step scheme.


\begin{table}[htbp]
\tabcolsep 0pt\caption{Convergence of Algorithm 3 for Example
\ref{ex_ns_1} with $h=1/128$ and $Re=5000$}\vspace*{-10pt}
\begin{center} \def\temptablewidth{\textwidth}
{\rule{\temptablewidth}{1pt}}
\begin{tabular*}{\temptablewidth}{@{\extracolsep{\fill}}llclc}
 $\Delta t$ & $\|\u-\u_h\|$ & order& $\|p-p_h\|$ &order\\
 \hline
 $0.2$ & 3.28203(-3) &- & 1.00222(-4)& -\\
 $0.1$ & 1.65607(-3) &0.9868& 4.79084(-5)&1.0648\\
 $0.1/2^1$ & 8.31889(-4) &0.9933& 2.35900(-5)&1.0221\\
 $0.1/2^2$& 4.16919(-4) &0.9966& 1.20411(-5)&0.9702 \\
\hline
\end{tabular*}
\end{center}\label{tab_ns_t1_1}
\end{table}

\begin{table}[htbp]
\tabcolsep 0pt\caption{Convergence of Algorithm 3 for Example
\ref{ex_ns_1} with $h=1/128$ and $Re=10000$}\vspace*{-10pt}
\begin{center} \def\temptablewidth{\textwidth}
{\rule{\temptablewidth}{1pt}}
\begin{tabular*}{\temptablewidth}{@{\extracolsep{\fill}}llclc}
 $\Delta t$ & $\|\u-\u_h\|$ & order& $\|p-p_h\|$ &order\\
 \hline
 $0.2$ & 3.28203(-3) & - & 9.98106(-5)&-\\
 $0.1$ & 1.65607(-3) &0.9872& 4.77006(-5)&1.0652\\
 $0.1/2^1$ & 8.31889(-4) &0.9935& 2.34866(-5)&1.0222\\
 $0.1/2^2$& 4.16919(-4) &0.9967& 1.19910(-5)&0.9699 \\
\hline
\end{tabular*}
\end{center}\label{tab_ns_t1_2}
\end{table}

%
\begin{table}[htbp]
\tabcolsep 0pt\caption{Convergence of Algorithm 3 for Example \ref{ex_ns_1}
with $\Delta t=10^{-6}$, $Re=5000$ and $T=0.2$}\vspace*{-10pt}
\begin{center} \def\temptablewidth{\textwidth}
{\rule{\temptablewidth}{1pt}}
\begin{tabular*}{\temptablewidth}{@{\extracolsep{\fill}}llclc}

 $h$ & $\|\u-\u_h\|$ & order& $\|p-p_h\|$ &order\\
\hline
 $1/4$ &  1.31468(-3) &-&   6.45683(-3)&-\\
 $1/8$ &  1.81020(-4)&2.8607& 1.61419(-3) &2.000016\\
 $1/16$ & 2.38018(-5) &2.9270&  4.03547(-3)&2.000002\\
 $1/32$& 3.00134(-6)  &2.9874 & 1.00887(-4)& 1.999996\\
 $1/48$& 8.71038(-7)  &3.0511& 4.48386(-5)&2.000004\\
\hline
\end{tabular*}
\end{center}\label{tab_ns_h1}
\end{table}

\begin{table}[htbp]
\tabcolsep 0pt\caption{Convergence of Algorithm 3 for Example \ref{ex_ns_1}
with $\Delta t=10^{-6}$, $Re=10000$ and $T=0.2$}\vspace*{-10pt}
\begin{center} \def\temptablewidth{\textwidth}
{\rule{\temptablewidth}{1pt}}
\begin{tabular*}{\temptablewidth}{@{\extracolsep{\fill}}llclc}

$h$ & $\|\u-\u_h\|$ & order& $\|p-p_h\|$ &order\\
\hline
 $1/4$ &  1.31577(-3) &-&   6.45686(-3)&-\\
 $1/8$ &  1.81589(-4)&2.8572& 1.61419(-3)&2.000016\\
 $1/16$ & 2.42194(-5) &2.9064&  4.03547(-4) &2.000002\\
 $1/32$& 3.20061(-6)  &2.9197 &1.00887(-3)& 1.999996\\
 $1/48$& 9.28064(-7)  &3.0533& 4.48386(-5)&2.000004\\
\hline
\end{tabular*}
\end{center}\label{tab_ns_h2}
\end{table}

\subsubsection{Stability improvement by the multistep scheme}\label{subsec_multistep_test_ns}
As shown in the previous subsection, the convergence of the
single-step scheme (Algorithm 3) for Example \ref{ex_ns_2} requires
a sufficiently small global time step size for a fixed mesh size
$h$.
In order to improve this severe restriction on time step size by the
single-step scheme, we now show how we can achieve the convergence
for large global time step size by the multistep scheme. For each
fixed $\Delta t=0.1/2^k$ ($k=-1, 0, 1, 2, \cdots$), we run the
multistep scheme with index $m=1, 2^1, 2^2, \cdots$ until we observe
the convergence of the scheme, and then record the corresponding
index $m$; see Tables \ref{tab_ns_t2_vm_Re_5000} and
\ref{tab_ns_t2_vm_Re_10000} for the recorded index $m$ corresponding
to each fixed $\Delta t$ and the resulting relative $L^2$-norm
errors of the approximate solutions for the velocity and pressure.

As we see from Table \ref{tab_ns_t2_vm_Re_5000}, when we take the global time step
$\Delta t =0.1$,
which is too large for the stability of  the explicit scheme involved in the convection step,
but we can still achieve the convergence of the multistep scheme
with index $m\ge 32$.
Tables \ref{tab_ns_t2_vm_Re_5000} and \ref{tab_ns_t2_vm_Re_10000}
have demonstrated that
though the single-step scheme does not converge for a fixed $\Delta t$,
the multistep scheme always converges when the index $m$ is appropriately large.
So we can conclude that if we take an appropriately large index $m$, say $m=30$,
the multistep scheme can be viewed as an unconditionally stable scheme.

Next we have tested the actual convergence orders of the multistep
scheme when the index $m$ is fixed at $m=64$. Tables
\ref{tab_ns_t2_m_64_Re_5000}-\ref{tab_ns_t2_m_64_Re_10000} have
showed the computational results for $Re=5000$  and $10000$ with
fixed $h=1/48$ and $1/64$ respectively. We can observe clearly the
optimal first order convergence for both velocity and pressure in
terms of the global time step size.

The last test we have carried out is to check how the multistep
scheme can improve the stability region quantitatively. For each
fixed mesh size $h$, we increase the index $m$ gradually and record
the largest global time step size $\Delta t$ (the critical time step
size $\Delta t_{crit}$ as we called earlier) that can ensure the
convergence of the entire algorithm. The results are shown in Table
\ref{tab_ns_t2_critical_time_step}, from which we can see that the
stability region is nearly doubled when the index $m$ of the
multistep scheme is doubled. So the multistep scheme can indeed
clearly and essentially enlarge the stability of the entire
algorithm.


\ss
We end this subsection with some concluding remarks  on convergence and
stability behaviors of the single-step and multistep schemes, based on our observations
from the numerical tests in this and previous subsections.
\begin{itemize}
  \item  The single-step scheme (Algorithm 3) is generally conditionally stable,
and requires sufficiently small time step size to ensure its
convergence with a fixed mesh and larger Reynolds number.
  \item The multistep scheme (Algorithm 4) can essentially relax the restriction
  of the time step size (see Tables \ref{tab_ns_t2_vm_Re_5000}, \ref{tab_ns_t2_vm_Re_10000}
  and \ref{tab_ns_t2_critical_time_step}), and behaves like a nearly unconditionally stable
  scheme.

\item Comparing the results in Tables
\ref{tab_ns_t2_vm_Re_5000}-\ref{tab_ns_t2_vm_Re_10000} with the ones
in Tables
\ref{tab_ns_t2_m_64_Re_5000}-\ref{tab_ns_t2_m_64_Re_10000}, we can
clearly see the stability and robustness of the multistep scheme
(Algorithm 4). For example, for the global time step size $\Delta
t=0.1/2^4$, the multistep scheme with a small index like $m=2$ and a
large index like $m=64$ provides about the same accuracies; see
Tables \ref{tab_ns_t2_vm_Re_10000} and
\ref{tab_ns_t2_m_64_Re_10000}.
\end{itemize}

\begin{table}[htbp]
\tabcolsep 0pt\caption{Stability of Algorithm 4 for Example
\ref{ex_ns_2} with index $m$ and fixed $h=1/48$, $Re=5000$
}\vspace*{-10pt}
\begin{center} \def\temptablewidth{\textwidth}
{\rule{\temptablewidth}{1pt}}
\begin{tabular*}{\temptablewidth}{@{\extracolsep{\fill}}lrll}

 $\Delta t$ &m & $\|\u-\u_h\|$ & $\|p-p_h\|$ \\
\hline
$0.1/2^4$&1& 1.61352(-3) &  9.22561(-3)\\
 $0.1/2^3$&4&  3.15065(-3) & 1.82083(-2)\\
 $0.1/2^2$&8&6.32378(-3) &  3.52747(-2) \\
 $0.1/2^1$ &16&  1.34098(-2) &6.64021(-2)\\
 $0.1$& 32&  2.70632(-2) & 1.18921(-1)\\
 $0.2$&64 & 5.60465(-2) &  1.97385(-1)\\
\hline
\end{tabular*}
\end{center}\label{tab_ns_t2_vm_Re_5000}
\end{table}

\begin{table}[htbp]
\tabcolsep 0pt\caption{Stability of Algorithm 4 for Example
\ref{ex_ns_2} with index $m$ and fixed $h=1/64$, $Re=10000$
}\vspace*{-10pt}
\begin{center} \def\temptablewidth{\textwidth}
{\rule{\temptablewidth}{1pt}}
\begin{tabular*}{\temptablewidth}{@{\extracolsep{\fill}}lrll}

 $\Delta t$ &m & $\|\u-\u_h\|$ & $\|p-p_h\|$ \\
\hline
$0.1/2^5$&1& 8.13177(-4)   &  4.65014(-3)\\
$0.1/2^4$&2& 1.61134(-3) &  9.24510(-3)\\
 $0.1/2^3$&4&  3.46105(-3)  & 1.82018(-2)\\
 $0.1/2^2$&16&6.37609(-3) &  3.52913(-2) \\
 $0.1/2^1$ &32&  1.29496(-2)  &6.64033(-2)\\
 $0.1$& 64&  2.67569(-2)  & 1.18926(-1)\\
 $0.2$&128& 5.67472(-2)&1.97454(-1) \\
\hline
\end{tabular*}
\end{center}\label{tab_ns_t2_vm_Re_10000}
\end{table}

\begin{table}[htbp]
\tabcolsep 0pt\caption{Convergence order of Algorithm 4 for Example
\ref{ex_ns_2} with fixed $h=1/48$, $Re=5000$ and fixed index
$m=64$}\vspace*{-10pt}
\begin{center} \def\temptablewidth{\textwidth}
{\rule{\temptablewidth}{1pt}}
\begin{tabular*}{\temptablewidth}{@{\extracolsep{\fill}}llclc}

 $\Delta t$ & $\|\u-\u_h\|$ & order& $\|p-p_h\|$ &order\\
  \hline
 $0.2$ & 5.60465(-2) &-& 1.97385(-1)&-\\
 $0.1$ & 2.64298(-2) &1.0845& 1.18888(-1)&0.7314\\
 $0.1/2^1$ & 1.28200(-2) &1.0438& 6.63946(-2)&0.8405\\
 $0.1/2^2$& 6.33073(-3)&1.0180& 3.52954(-2) &0.9116\\
 $0.1/2^3$& 3.17694(-3) &0.9947& 1.82354(-2)&0.9527\\
 $0.1/2^4$& 1.60997(-3) &0.9806& 9.27332(-3)&0.9756\\
 $0.1/2^5$& 8.20838(-4) &0.9719& 4.67745(-3)&0.9874\\
\hline
\end{tabular*}
\end{center}\label{tab_ns_t2_m_64_Re_5000}
\end{table}

%
\begin{table}[htbp]
\tabcolsep 0pt\caption{Convergence order of Algorithm 4 for Example
\ref{ex_ns_2} with fixed $h=1/64$, $Re=10000$ and fixed index
$m=64$}\vspace*{-10pt}
\begin{center} \def\temptablewidth{\textwidth}
{\rule{\temptablewidth}{1pt}}
\begin{tabular*}{\temptablewidth}{@{\extracolsep{\fill}}llclc}

 $\Delta t$ & $\|\u-\u_h\|$ & order& $\|p-p_h\|$ &order\\
  \hline

 $0.1$ & 2.67569(-2)  &-& 1.18926(-1) &-\\
 $0.1/2^1$ & 1.29252(-2) &1.0497& 6.64054(-2)&0.8407\\
 $0.1/2^2$& 6.35955(-3)&1.0232& 3.52978(-2) &0.9117\\
 $0.1/2^3$& 3.17773(-3) &1.0009& 1.82330(-2)&0.9530\\
 $0.1/2^4$& 1.60477(-3) &0.9856& 9.27095(-3)&0.9758\\
 $0.1/2^5$& 8.16528(-4) &0.9748& 4.67451(-3)&0.9879\\
\hline
\end{tabular*}
\end{center}\label{tab_ns_t2_m_64_Re_10000}
\end{table}

\begin{table}[htbp]
\tabcolsep 0pt\caption{Critical global time step size $\Delta
t_{crit}$ of Algorithm 4 for Example \ref{ex_ns_2} in terms of index
$m$}\vspace*{-10pt}
\begin{center} \def\temptablewidth{\textwidth}
{\rule{\temptablewidth}{1pt}}
\begin{tabular*}{\temptablewidth}{@{\extracolsep{\fill}}lcccccc}

 $m$ & 1 & 5& 10& 20& 40&80  \\
\hline
 $Re=10000, h=1/64$\\
 $\Delta t_{crit}$&
0.0039&0.018&0.024&0.048&0.089&0.18\\
\hline
\end{tabular*}
\end{center}\label{tab_ns_t2_critical_time_step}
\end{table}


\subsubsection{The lid-driven cavity flow}\label{subsec_cavity_flow}
As our final numerical example we test a popular benchmark problem,
i.e., the lid-driven cavity flow problem, where the fluid is
enclosed in a unit square box, with an imposed velocity of unity in
the horizontal direction on the top boundary, and a no-slip
condition on the remaining walls.   We shall compare our results
with three benchmark results: Ghia et al.\,\cite{GGS1982} with
$h=1/128$ for Reynolds numbers $Re=100, 400,1000$ and $3200$; Erturk
et al.\,\cite{ECG2005} with $h=1/128$ for Reynolds number $Re=1000$;
Botella et al.\,\cite{BP1998} for the Reynolds number $Re=1000$.

In all our computations for this example,  we use the uniform mesh
of size $h=1/128$ and the Taylor-Hood elements
(\ref{Taylor-Hood_element}), and have tested the cases with Reynolds
numbers $Re=100,400,1000$ and $3200$, and the global time step size
$\Delta t=0.004$. The stoping condition for time advancing, which is
considered as the criterion of capturing the steady state solution,
is chosen as
$$\frac{\|\u_h^{n+1}-\u_h^{n}\|}{\|\u_h^{n+1}\|}\le 10^{-5}\, ,$$
where $\u_h^{n}$ is the finite element solution at time $t=t_n$.
We have observed from our numerical results that
the single-step scheme (Algorithm 3) works when the Reynolds number
is relatively small, e.g., $Re=100,400$ and $1000$, but it is unstable
when $Re$ is large, e.g., $Re= 3200$.
But the multistep scheme may still work for larger  Reynolds number,
e.g., $Re\ge 3200$.

\begin{figure}[h] 
\centering
\includegraphics[width=6cm]{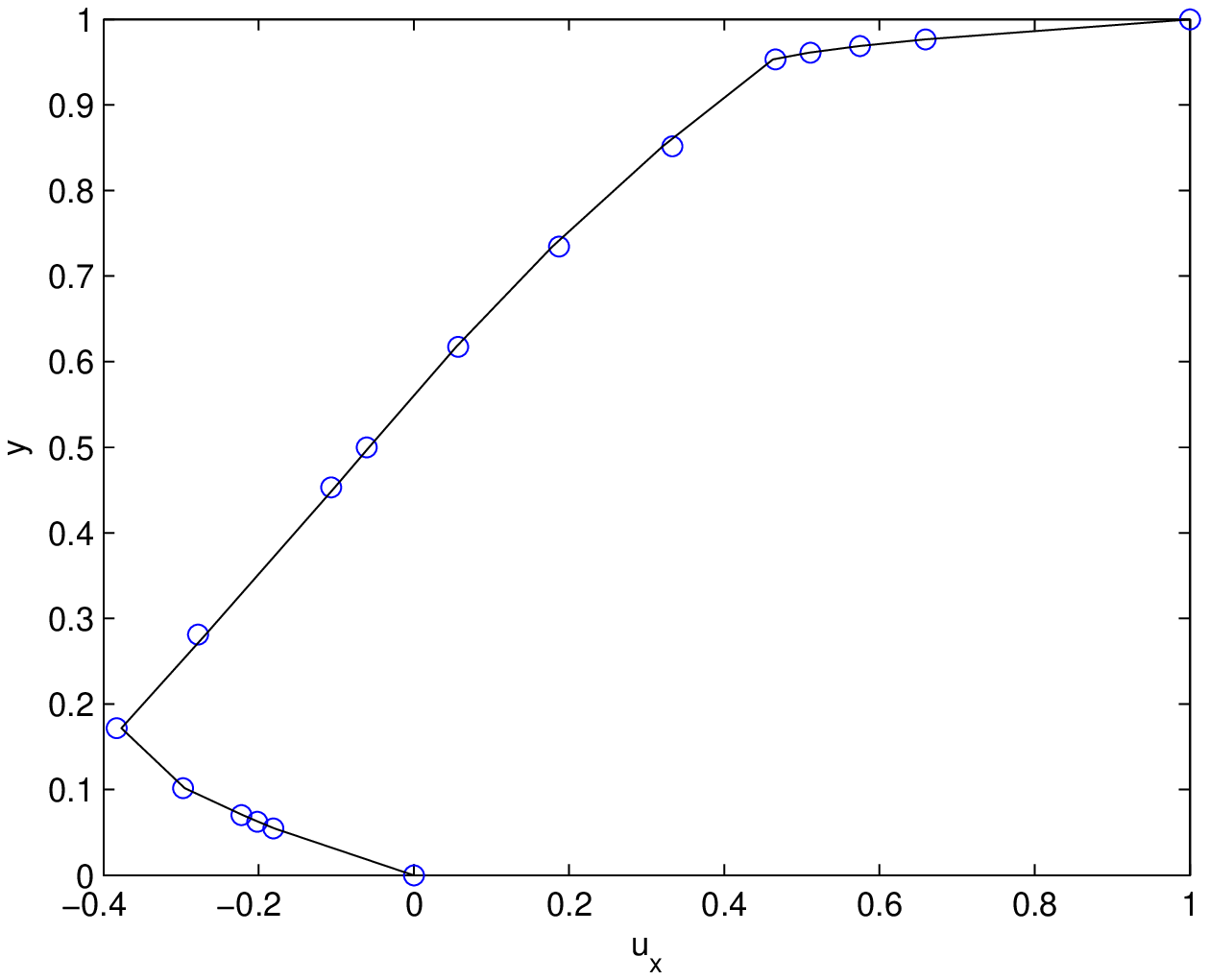}
\includegraphics[width=6cm]{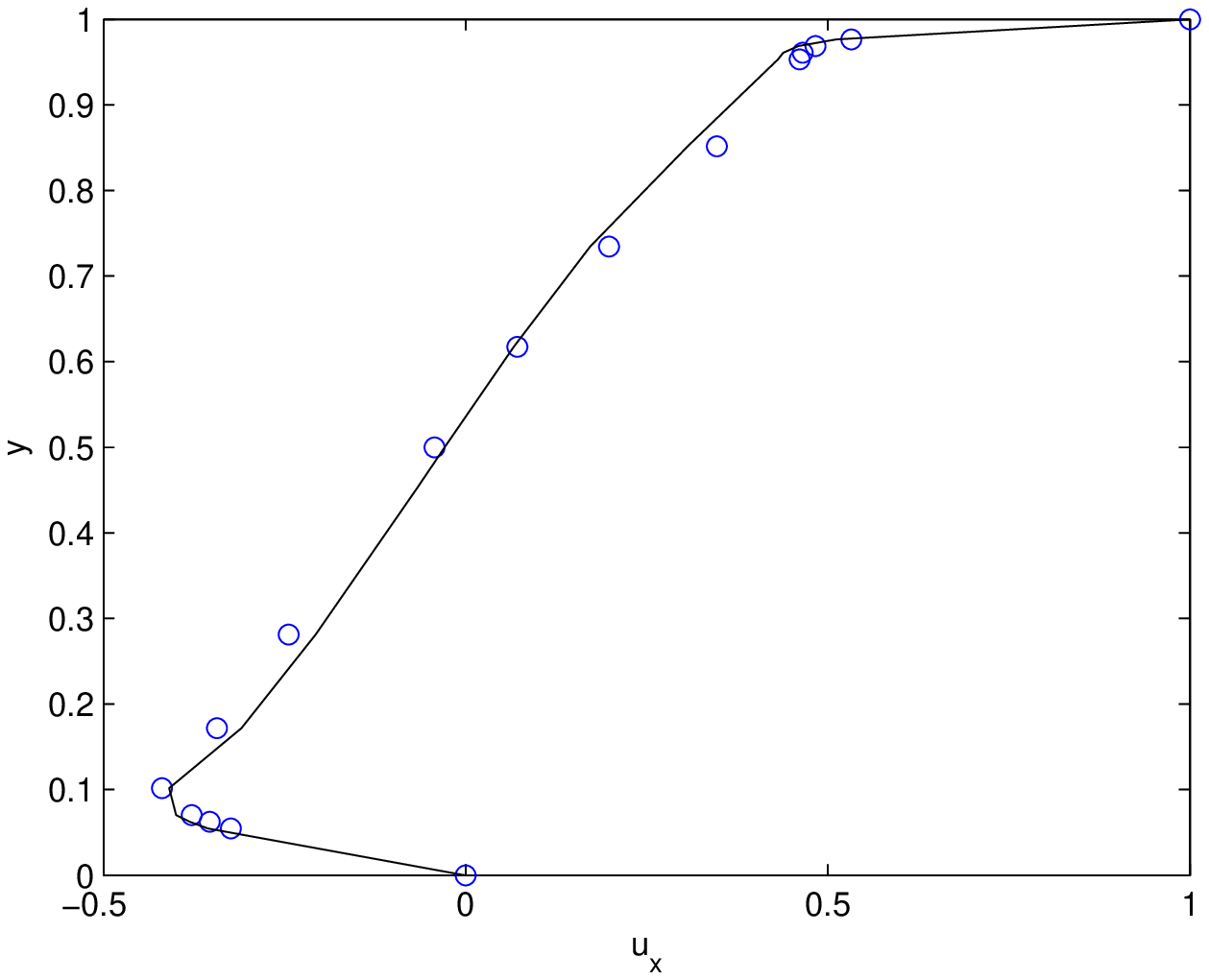}\\
{\centerline{(a) $Re=1000$\qquad\qquad\qquad\qquad\qquad\qquad (b)
$Re=3200$}}
 \caption{Velocity ($u_x$) profiles along the vertical line passing through the geometric center
 of the cavity. Black solid lines: (a) single-step scheme, (b) multistep scheme with index $m=2$; Blue circle
lines: Ghia et al.\,\cite{GGS1982} }\label{fig_cavity_u_profile}
\end{figure}

\begin{figure}[h] 
\centering
\includegraphics[width=6cm]{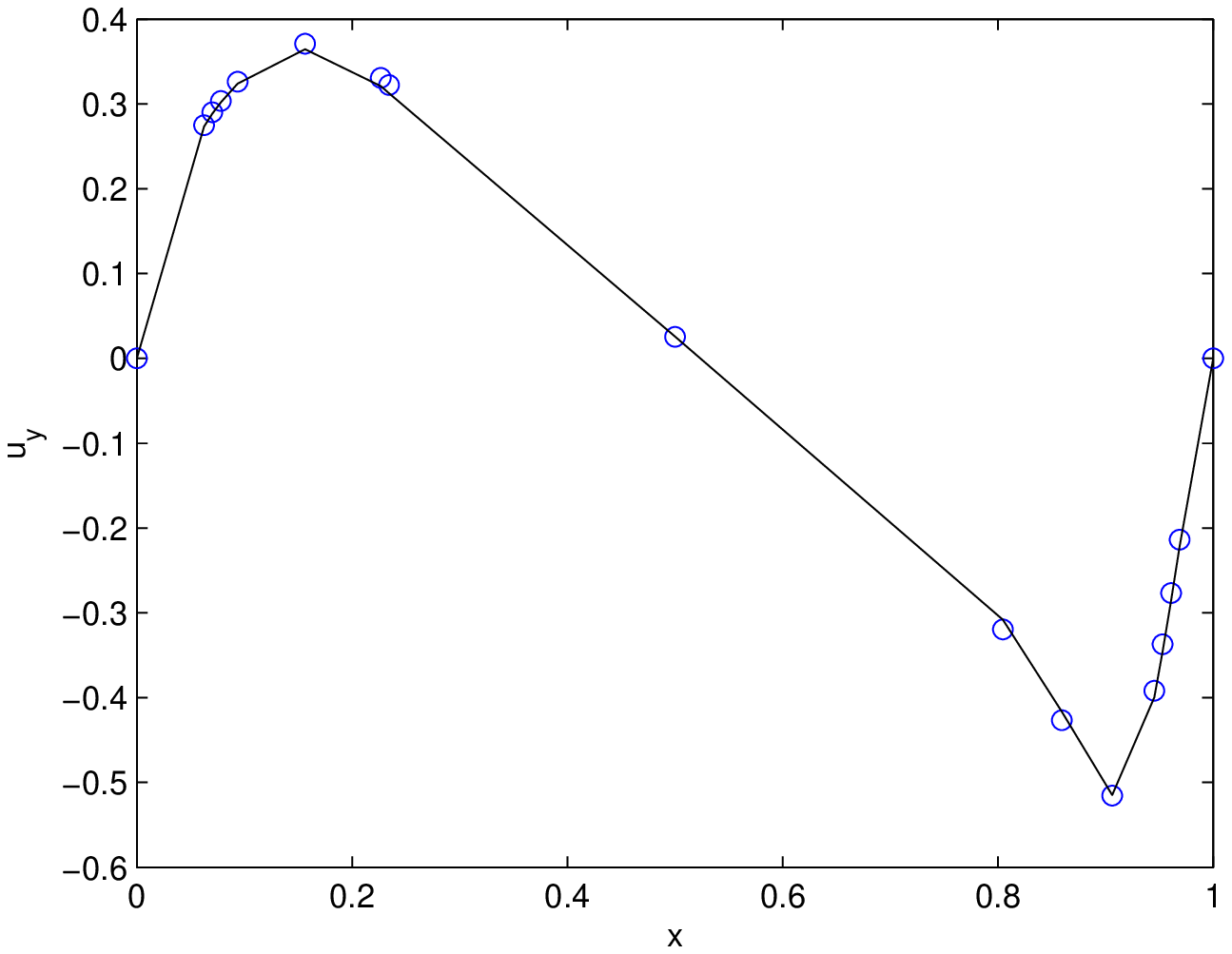}
\includegraphics[width=6cm]{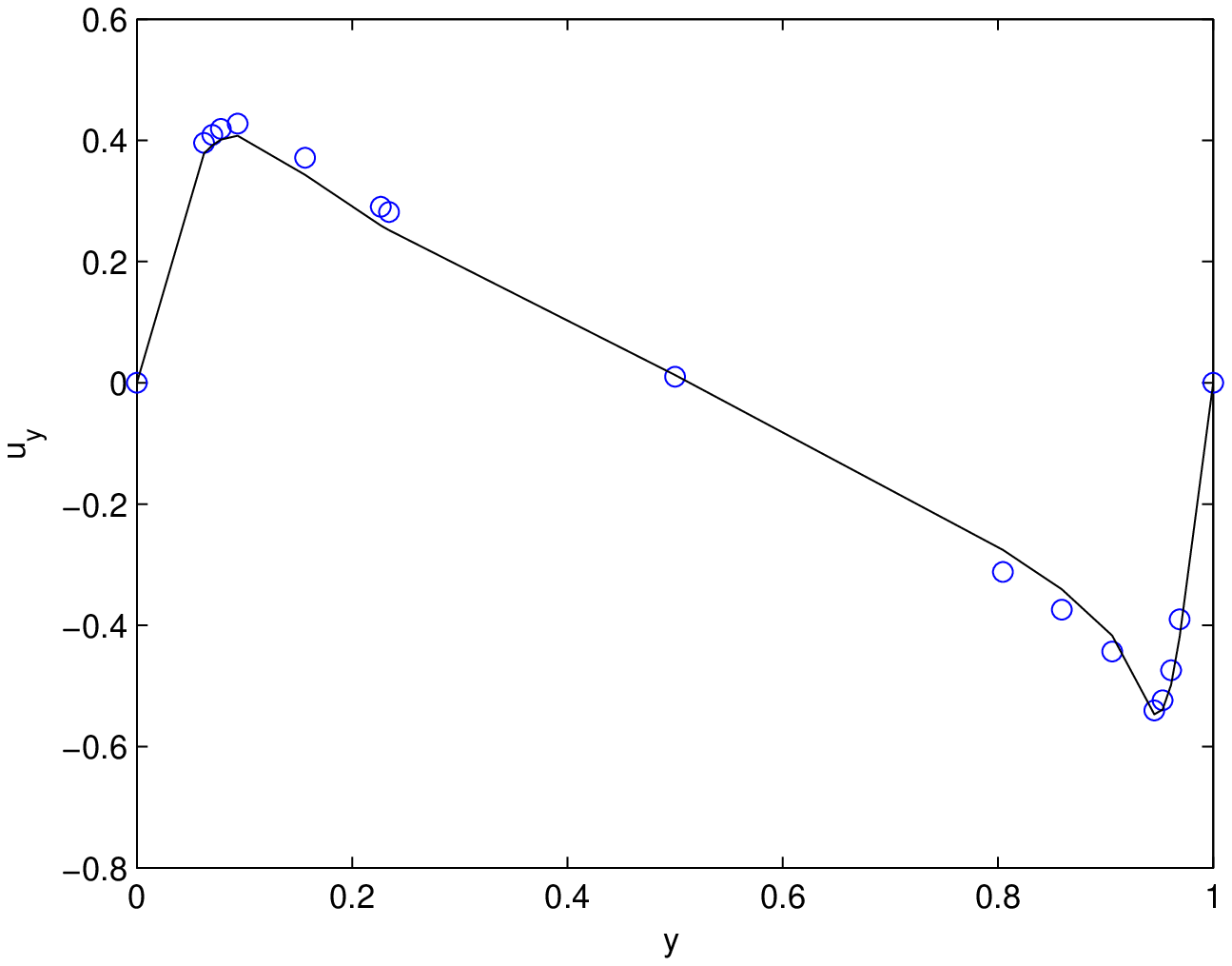}\\
{\centerline{(a)$Re=1000$\qquad\qquad\qquad\qquad\qquad\qquad
(b)$Re=3200$}}
 \caption{Velocity ($u_y$) profiles along the horizontal line passing through the geometric center
 of the cavity. Black solid lines: (a) single-step scheme, (b) multistep scheme with index $m=2$; Blue circle
lines: Ghia et al.\,\cite{GGS1982}}\label{fig_cavity_v_profile}
\end{figure}

\begin{figure}[htbp]
\centering
\includegraphics[width=6cm]{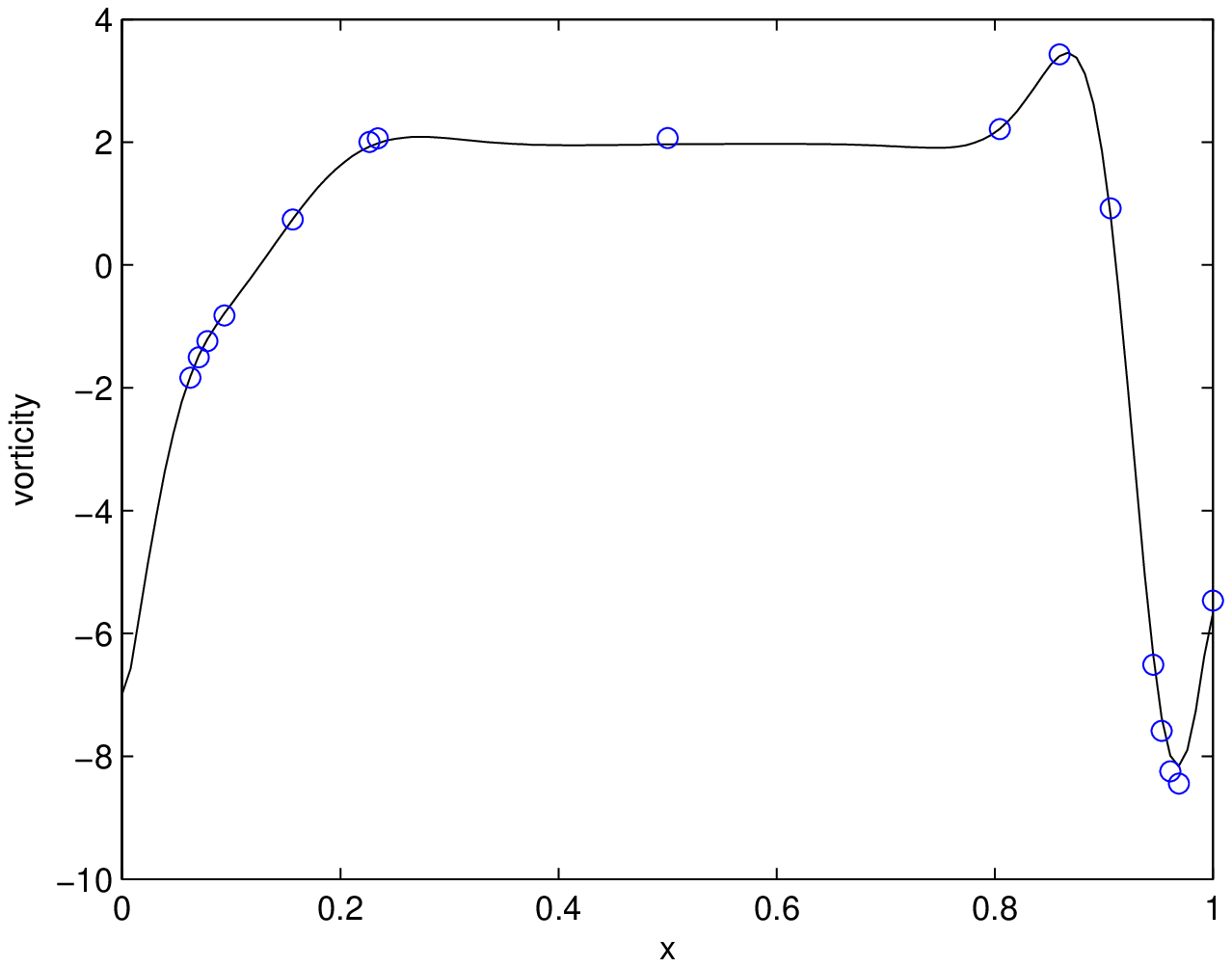}
\includegraphics[width=6cm]{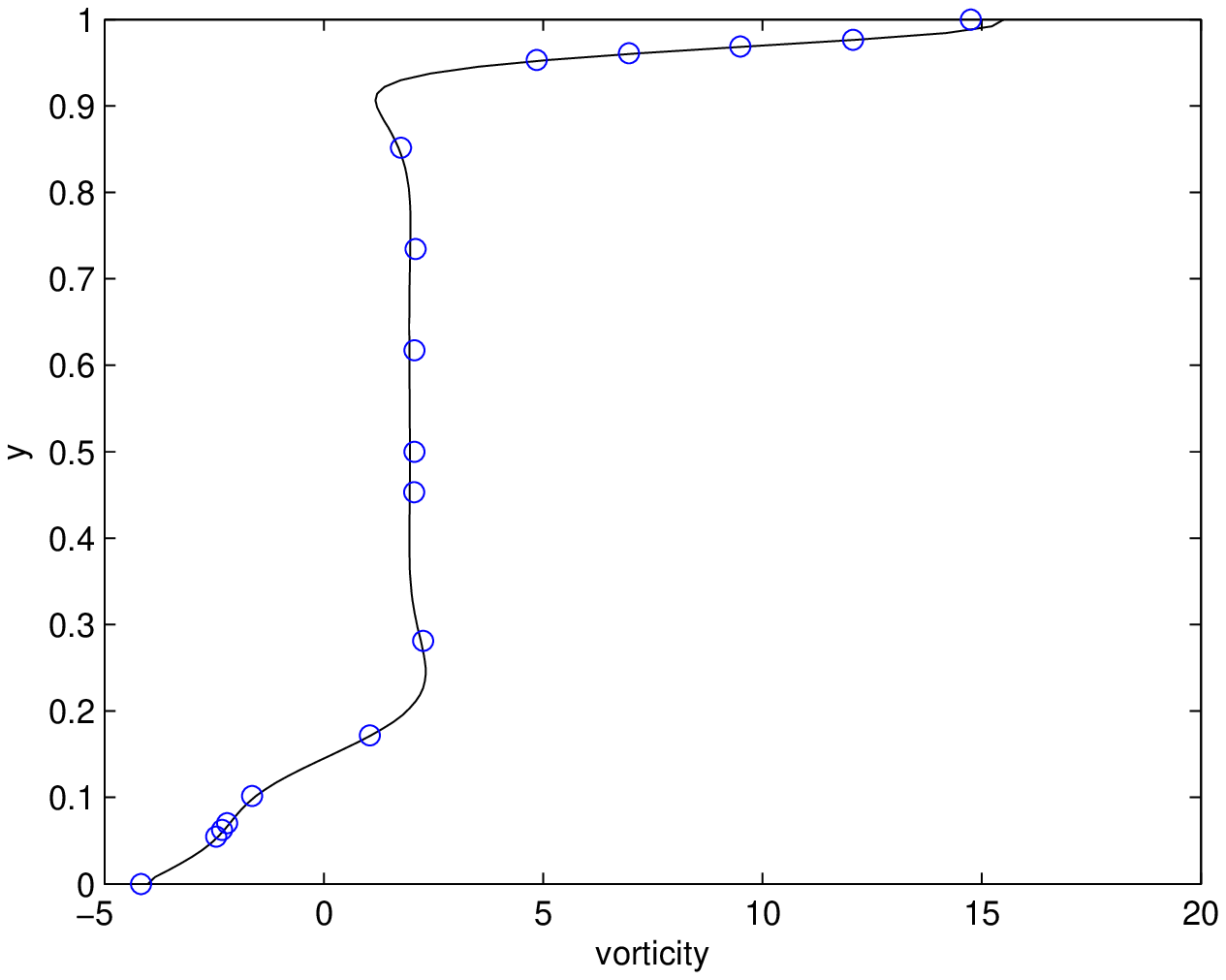}\\
 \caption{Vorticity values along the vertical line $x=0.5$ (left) and the horizontal line $y=0.5$
(right) passing through the geometric center of the cavity with
$Re=1000$. Black solid lines: single-step scheme; Blue circle lines:
Botella et al.\,\cite{BP1998}}\label{fig_cavity_vorticity_profile}
\end{figure}

Tables \ref{tab_cavity_Re_1000}-\ref{tab_cavity_Re_3200} present the
streamfunction values and the locations of the primary and secondary
vortices for various Reynolds numbers. Figures
\ref{fig_cavity_u_profile}, \ref{fig_cavity_v_profile} and
\ref{fig_cavity_vorticity_profile} show the computed velocity
components and vorticity profiles along the horizonal and vertical
lines compared with the results of Ghia et al.\,\cite{GGS1982} and
Botella et al.\,\cite{BP1998}. As one can see that the results by
the new schemes confirm very well with the ones by the benchmark schemes.

%

\begin{table}[htbp]
\tabcolsep 0pt\caption{Streamfunction values $\Psi_{min}$,
$\Psi_{max}$ and locations  of the primary and secondary vortices
}\vspace*{-10pt}
\begin{center} \def\temptablewidth{\textwidth}
{\rule{\temptablewidth}{1pt}}
\begin{tabular*}{\temptablewidth}{@{\extracolsep{\fill}}lcccccl}
 Vortex &property &Re=1000 &Re=1000&Re=1000 \\
 & &Single-step scheme &Ghia  et al.\,\cite{GGS1982}&Erturk et al.\,\cite{ECG2005}\\
 \hline
Primary&$\Psi_{min}$&-0.114722&-0.117929&-0.118781 \\
&Location (x, y)&(0.5313, 0.5625)&(0.5313, 0.5625)&(0.5300, 0.5650)\\
First BL&$\Psi_{max}$&2.12504E-4&2.31129E-4&2.3261E-4\\
&Location (x, y)&(0.0781, 0.0781)&(0.0859, 0.0781)&(0.0833, 0.0783)\\
First BR&$\Psi_{max}$&1.67313E-3&1.75102E-3&1.7281E-3\\
&Location (x, y)&(0.8672, 0.1094)&(0.8594,0.1094)&(0.8633, 0.1117)\\
Second BR&$\Psi_{min}$&-4.815059E-8&-9.31929E-8&5.4962E-8\\
&Location (x, y)&(0.9922, 0.0078)&(0.9922, 0.0078)&(0.9917, 0.0067)\\
\hline
\end{tabular*}
\end{center}\label{tab_cavity_Re_1000}
\end{table}

\begin{table}[htbp]
\tabcolsep 0pt\caption{ Streamfunction values $\Psi_{min}$,
$\Psi_{max}$ and locations  of the primary and secondary vortices
}\vspace*{-10pt}
\begin{center} \def\temptablewidth{\textwidth}
{\rule{\temptablewidth}{1pt}}
\begin{tabular*}{\temptablewidth}{@{\extracolsep{\fill}}lcccl}
 Number &property  &Re=3200&Re=3200\\
 & &Multistep scheme with index $m=2$ &Ghia et al.\,\cite{GGS1982}\\
 \hline
Primary&$\Psi_{min}$&-0.109962&-0.120377 \\
&Location, x, y&(0.5156,0.5391)&(0.5165,0.5469)\\

First T&$\Psi_{max}$&5.759079E-4&7.27682E-4\\
&Location (x, y)&(0.0469,0.8984)&(0.0547,0.8984)\\
First BL&$\Psi_{max}$&1.09512E-3&9.7823E-4\\
&Location (x, y)&(0.0781,0.1250)&(0.0859,0.1094)\\
First BR&$\Psi_{max}$&2.70425E-3&3.13955E-3\\
&Location (x, y)&(0.8281,0.0859)&(0.8125,0.0859)\\

Second BL&$\Psi_{min}$&-1.04040E-8&-6.33001E-8\\
&Location (x, y)&(0.0078,0.0078)&(0.0078,0.0078)\\
Second BR&$\Psi_{min}$&-1.36461E-7&-2.51648E-7\\
&Location (x, y)&(0.9844,0.0078)&(0.9844,0.0078)\\
\hline
\end{tabular*}
\end{center}\label{tab_cavity_Re_3200}
\end{table}

\section{Concluding remarks}\label{sec_conclus}
We have proposed a new splitting method for solving
time-dependent convection-dominated diffusion problems.
A pure convection problem and a pure
diffusion problem are solved successively at each iteration of the
method. Explicit schemes are proposed for the time discretization of
the convective problem. The explicitness of the scheme may cause a
severe restriction on the time step size, which can be essentially
improved by an explicit multistep scheme with smaller time step
sizes so that the resulting method behaves like an unconditionally
stable method. The diffusion problem involved at each iteration is
always self-adjoint and coercive so that it can be solved
efficiently using many existing optimal preconditioned iterative
solvers. The optimal convergence orders have been confirmed by
several numerical examples with smooth solutions.
The schemes are then extended for the Navier-Stokes equations, where
the nonlinearity is resolved by a linear explicit multistep scheme
at the convection step, while only a  generalized Stokes problem is
needed to solve at the diffusion step and the major stiffness matrix
stays invariant in the time marching process. Numerical simulations
are presented to demonstrate the stability, convergence and
performance of the single-step and multistep variants of the new
schemes. The effectiveness and robustness of the new schemes are
finally well demonstrated by the benchmark lid-driven cavity flow
problem. The newly proposed schemes are all free from tuning some stabilization
parameters as the most existing schemes require for the convection-dominated
diffusion problems.
Finally we note that the proposed fully discrete schemes are only first order in time,
and we are currently investigating the potential schemes which
have second order temporal accuracy.

\end{document}